\documentclass[twocolumn,tbtags,usenames,dvipsnames]{aastex62}

\newcommand{\del}{\Vec{\mathbf{\nabla}}}

\newcommand{\bnabla}{{\mbox{\boldmath$\nabla$}}}

\newcommand{\mbfB}{\mathbf{B}}

\newcommand{\mbfu}{\mathbf{u}}

\newcommand{\mbfvA}{\mathbf{v}_\mathrm{A}}

\newcommand{\tinjo}{\tau_\mathrm{inj,0}}
\newcommand{\tlosso}{\mathrm{\tau_{C,0}}}

\newcommand{\lr}[1]{\left(#1\right)}
\newcommand{\unit}{\rm \,}

\received{\today}

\submitjournal{ApJ}

\shorttitle{Cosmic Ray Eddington}
\shortauthors{Heintz \& Zweibel}

\usepackage[tbtags]{amsmath}
\usepackage{natbib}
\usepackage{xcolor}
\usepackage{bm}
\usepackage{graphicx}                
\usepackage{amssymb}
\usepackage{subfigure}
\usepackage{footnote}
\usepackage{dblfnote}

\begin{document}

\title{Galaxies at a Cosmic-Ray Eddington Limit}

\correspondingauthor{Evan Heintz}
\email{eheintz@wisc.edu, zweibel@astro.wisc.edu}

\author{Evan Heintz}
\affil{Department of Physics, University of Wisconsin - Madison, 1150 University Ave., Madison, WI 53706, USA}

\author{Ellen G. Zweibel}
\affil{Department of Physics, University of Wisconsin - Madison, 1150 University Ave., Madison, WI 53706, USA}
\affil{Department of Astronomy, University of Wisconsin - Madison, 475 North Charter St., Madison, WI 53706, USA}


\begin{abstract}
Cosmic rays have been shown to be extremely important in the dynamics of diffuse gas in galaxies, helping to maintain hydrostatic equilibrium, and serving as a regulating force in star formation. In this paper, we address the influence of cosmic rays on galaxies by re-examining the theory of a cosmic ray Eddington limit, first proposed by \cite{socrateseddington2008} and elaborated upon by \cite{CrockerEddington2021Part1} and \cite{HuangDavis2022}. A cosmic ray Eddington limit represents a maximum cosmic ray energy density above which the interstellar gas cannot be in hydrostatic equilibrium, resulting in a wind. In this paper, we continue to explore the idea of a cosmic ray Eddington limit by introducing a general framework that accounts for the circumgalactic environment and applying it to five galaxies that we believe to be a good representative sample of the star forming galaxy population, using different cosmic ray transport models to determine what gives each galaxy the best chance to reach this limit. We show that while an Eddington limit for cosmic rays does exist, for our five galaxies, the limit either falls at star formation rates that are much larger or gas densities that are much lower than each galaxy's measured values. This suggests that cosmic ray pressure is not the main factor limiting the luminosity of starburst galaxies.
\end{abstract}

\section{Introduction}
\label{sec:intro}

Cosmic rays play an important role in many different phenomena in our Universe. In recent decades, one of the many areas of study concerning them has been star formation feedback. Many papers have shown the role of cosmic rays in regulating star formation by puffing up the galactic gas layers and helping launch a galactic wind \citep{breitschwerdtwinds1991,everettwinds2008,uhligwinds2012, boothwinds2013,hanaszwinds2013,SalemCROutflows2014, girichidiswinds2016,simpsonWinds2016,ruszkowskiwinds2017,MaoCRWind2018,HopkinsCRGalaxyForm2020,BustardLMCWinds2020}. In these works, the effects of cosmic rays on the evolution of the galaxy is largely dependent on the transport model used to describe them. Most models show that if the cosmic rays diffuse so weakly that they are effectively locked to the thermal gas (we will refer to this as transport by advection), they will puff up the galactic disk and quench star formation. But if the cosmic rays diffuse or stream along magnetic field lines, they are more effective in driving a galactic wind and limit star formation through that process, although less than in the advection case.

However, many of these papers only show that cosmic rays aid in driving galactic winds by supplementing the thermal pressure. They do not show that cosmic rays alone can drive galactic winds. The pioneering paper by \cite{socrateseddington2008} aimed to answer the question of whether a purely cosmic-ray driven wind was possible. In their paper, they laid out the theoretical framework for what they called a cosmic-ray Eddington limit, assuming that cosmic rays were fully supporting the gas against the gravitational field of an isothermal sphere. In this model, the idea is that a system could have such vigorous star formation that the resulting cosmic ray luminosity would be large enough to break hydrostatic equilibrium and blow out the interstellar medium by launching a wind. In other words, \cite{socrateseddington2008} claimed that there is a limiting star formation rate above which hydrostatic equilibrium cannot be maintained due to the cosmic ray pressure. Because of the close connection between the star formation rate and the cosmic ray production rate and the relatively short timescale on which blowout would occur, this would set an upper limit on the luminosity of starburst galaxies.

\cite{CrockerEddington2021Part1} and \cite{CrockerEddington2021Part2} recently reexamined the idea of a cosmic ray Eddington limit. In their paper, they assumed a horizontal magnetic field and took the gravitational potential to be that of a one-dimensional infinite slab, which accurately portrays the gravitational field close to the galactic disk but results in  constant gravitational acceleration far away from the disk. They found that in an Eddington type model, there exists a stability limit in galaxies above which hydrostatic equilibrium can no longer be satisfied. Generally, they found that galaxies with high gas surface densities and large star formation rates are less likely to be near this stability limit due to the large amounts of hadronic losses that occur in these systems. In related work, \cite{HuangDavis2022} examined the initial stages of wind launching by a fixed flux of cosmic rays at the base of an atmosphere with a vertical magnetic field along which cosmic rays could diffuse or stream.  This study showed that
the launch mechanism is quite robust under a range of models for cosmic ray transport and gas physics.

In this paper, we step back from the verisimilitude of the  \cite{CrockerEddington2021Part1} and \cite{HuangDavis2022} local 1D models
in order to focus on the global effects introduced by a bounded galactic gravitational potential and confining pressure of a circumgalactic medium.
A potential that allows $g$, the gravitational acceleration, to fall off with distance will make it more likely that galaxies reach the Eddington limit. In this work, we formulate our problem for a general gravitational potential and then substitute in specific examples that fall off with distance such as for a finite mass source, allowing for a system where the gas can escape the galaxy's gravity.

The cosmic ray transport model that best leads to reaching the cosmic ray Eddington limit is something else we wish to elaborate on.  \cite{socrateseddington2008} assumed a model that included cosmic ray streaming with a random walk component. \cite{CrockerEddington2021Part1} extended this and compared three different models of cosmic ray transport: streaming with random walk along field lines, scattering off extrinsic turbulence, and constant diffusion. \cite{HuangDavis2022} also considered streaming and diffusion. We will follow a similar procedure as \cite{CrockerEddington2021Part1} and \cite{HuangDavis2022}  and compare different transport models throughout our analysis (see \S\ref{subsec:CRTransport} for our choices).

To determine if a wind is likely to be blown out by the systems we are looking at, we will follow the arguments presented by \cite{ParkerSolarWind1958a} in regards to the solar wind. In that article, Parker assumed the solar corona to be isothermal, spherically symmetric, and in hydrostatic equilibrium. Under these assumptions, the asymptotic pressure $P(r)$ approaches an asymptotic value orders of magnitude larger than the interstellar pressure, so Parker concluded that hydrostatic equilibrium is impossible and the corona must be flowing outward.

For our models, we will usually do the same as \cite{ParkerSolarWind1958a} and derive an asymptotic cosmic-ray pressure from hydrostatic equilibrium assuming that cosmic-ray pressure is the only pressure supporting the system against gravity. We can then compare that value to the base value of the surrounding circumgalactic medium (CGM). If the cosmic ray pressure is determined to be larger than the pressure of the CGM, then we must conclude that hydrostatic equilibrium has been broken and a wind would be launched. We describe our procedure for non-asymptotic systems in \S \ref{subsec:numResults}.

In \S\ref{sec:analytical}, we outline the basic setup of our problem. In \S\ref{subsec:hydroEq}, we establish our equations for hydrostatic equilibrium, along with the galactic potential we will be using throughout our analysis. We continue with outlining the cosmic ray transport models we will be using in \S\ref{subsec:CRTransport} and describe how we implement sources and collisional losses in \S\ref{subsec:setupSC}. In \S \ref{subsec:transportModels}, we derive analytical forms for the gas density and cosmic ray transport for different models of cosmic ray transport without sources and collisions. In \S\ref{subsec:parameters}, we describe the five different galaxies that we will use to model the cosmic ray Eddington limit and then insert their parameters into our analytical solutions in \S \ref{subsec:anaResults}. We then consider the joint effects of transport, sources and collisions in \S\ref{subsec:numResults} and solve the equations numerically for the five galaxies of our analysis. Discussion of these results and a comparison between the analytical and numerical solutions can be found in \S\ref{sec:discuss}, along with our final conclusions. Details on the nondimensionalization of the equations and how we determined the parameters for the galaxies in our studies are in Appendices \ref{sec:nondimens} and \ref{sec:galacticparam}.

\section{Setup of the Problem}
\label{sec:analytical}
For this analysis, we will assume that we have a purely CR-supported atmosphere in a gravitational potential that arises from a spherically symmetric mass distribution at the center of the galaxy, along with a dark matter halo. This spherical symmetry will lend itself to using potentials that fall off with distance. Within the central mass distribution, which we envision as a region of intense star formation, we will assume a balance between cosmic ray sources and cosmic ray losses, which include both collisions and diffusive transport These conditions will then provide us with initial conditions at the edge of our mass distribution, many of which we will treat as constant parameters for each galaxy. There is no reason in principle why the central starburst region and the central mass distribution should have the same radius, but for simplicity we assume this to be the case.

For a system to reach the Eddington limit, one of two requirements must be met. It must either have vigorous enough star formation that enough cosmic rays are created to blow away the surrounding medium or it must have a low enough gas density that the cosmic ray pressure supplied by the galaxy can surpass the conditions for hydrostatic equilibrium. Therefore, we will specifically focus on the values of $\rho_\mathrm{0}$ and $U_\mathrm{c0}$, the gas density and the cosmic ray energy density at the edge of the central mass distribution, respectively. We will treat these as the two boundary conditions with which we solve equations later and we will often vary over them to find the values at which a system reaches the cosmic ray Eddington limit. 

A system with asymptotic pressure greater than the CGM pressure is certainly super-Eddington, but even if this condition is not met, a cosmic ray supported hydrostatic atmosphere is subject to at least two minimal reality checks. One is that radius at which $P_c=P_\mathrm{CGM}$, which we term the confinement radius $R_{conf}$, should be reasonable.  The second is that the mass confined:
\begin{equation}\label{eq:Mconf}
    M_{conf} \equiv 4\pi \int_R^{R_{conf}} \rho(r) r^2 dr
\end{equation}
should be a reasonable interstellar mass.

\subsection{Hydrostatic Equilibrium}
\label{subsec:hydroEq}
We will assume that our system starts in hydrostatic equilibrium and then will aim to determine if this equilibrium is eventually broken by a large enough cosmic ray pressure. Since we are assuming a system where the cosmic-rays alone are supporting the gas against gravity, our equilibrium condition  will be:
\begin{equation}
    \frac{dP_\mathrm{c}}{dr} = -\rho\frac{d\Phi}{dr}
    \label{eq:hydroequil}
\end{equation}
where $\Phi$ is the gravitational potential, $P_\mathrm{c}$ is the cosmic ray pressure, $\rho$ is the gas density, and $r$ is the distance from the center of the system.

Although in many cases the solution of eqn. (\ref{eq:hydroequil}) can be written in terms of a general $\Phi$, for quantitative results in this paper we will take a $\Phi$ that has a dark matter halo component in addition to the component from a spherical mass distribution at the center of the galaxy. From \cite{HernquistPotential1990}, the potential is:
\begin{equation}
    \Phi(r) = -\frac{GM_h}{(r+a)} - \frac{GM_\mathrm{c}}{r}
    \label{eq:halomasspot}
\end{equation}
where $M_\mathrm{c}$ is the core mass in the center, $M_h$ is the halo mass and $a$ is a scale length determined by a semilog plot between the two points ($6$ kpc, $10^{11} M_{\odot})$ and ($25$ kpc, $10^{13} M_{\odot}$). 

\subsection{Cosmic Ray Transport}
\label{subsec:CRTransport}
For this work, we will be analyzing the effect that different cosmic ray transport models have on the Eddington limit. We will assume transport is controlled by two processes: streaming at the Alfven speed $v_A$, and diffusion. 
The general steady state cosmic ray transport equation, taking into account both processes, is \cite{breitschwerdtwinds1991}:
\begin{equation}
\begin{split}
   \bnabla\cdot\left[\gamma\left(\mbfu +\mbfvA\right) P_c- \kappa \bnabla P_c\right]=\\ 
    (\gamma - 1)\left(\mbfu+\mbfvA\right)\cdot \bnabla P_c +
    (\gamma - 1)Q
    \label{eq:fulltransport}
    \end{split}
\end{equation}
where $\gamma$ is the adiabatic index of the cosmic rays (canonically taken to be 4/3), $\mathbf{u}$ is the velocity of the gas, $\mathbf{v_\mathrm{A}} = \mathbf{B}/(4\pi\rho)^{1/2}$ is the Alfv\'{e}n speed, $\kappa$ is the diffusion tensor, and $Q$ represents both source and losses from collisions in the system. One  subtlety to keep in mind is that $v_A$ is to be computed with respect to the \texttt{plasma} density, not the total gas density. That is because the frequencies of the waves which scatter the cosmic rays are generally higher than the rates at which ions and neutrals collide, so the waves propagate in the plasma component alone \citep{KulsrudSelfConfine1969}. For our analysis, we will look at the limit where only one transport process is present and determine the forms of $P_\mathrm{c}$ in that limit. This will allow us to determine  how each transport process affects the ability of a galaxy to reach the cosmic ray Eddington limit. We go into more detail on these "pure" transport processes in \S\ref{subsec:transportModels}.

In the self-confinement model \citep{KulsrudSelfConfine1969}, if the cosmic ray bulk velocity, $v_D$, is larger than the Alfv\'{e}n speed, $v_\mathrm{A}$, the the cosmic rays will excite Alfv\'{e}n waves through the streaming instability. The cosmic rays will be confined by these same waves, scattering off of them 
until they reach isotropy in the wave frame. In a steady state, these waves will transfer energy and momentum to the surrounding gas \citep{zweibelreview2017}.

In the model with purely cosmic ray streaming, we drop the $Q$ term, 
assume diffusion is negligible, and set $\mathbf{u} = 0$ so that eqn. (\ref{eq:fulltransport}) becomes:
\begin{equation}\label{eq:streamingonly1}
     \bnabla\cdot\lr{\gamma\mathbf{v_\mathrm{A}}P_\mathrm{c}} = (\gamma - 1)\mathbf{v_\mathrm{A}}\cdot \bnabla P_\mathrm{c}.
\end{equation}
We can then use $\bnabla\cdot\mbfB\equiv 0$ to write $\bnabla \cdot \mathbf{v_\mathrm{A}} = - \left(\mbfvA \cdot \bnabla\rho \right)/ (2\rho)$ and reduce eqn. (\ref{eq:streamingonly1}) to the form: 
\begin{equation}
   - \gamma P_\mathrm{c} \frac{\mathbf{v_\mathrm{A}}}{2\rho}\cdot \bnabla \rho + \mathbf{v_\mathrm{A}} \cdot \bnabla P_\mathrm{c} = 0
\end{equation}
which when integrated leads to the polytropic relationship $P_c/\rho^{\gamma/2}$, meaning this term is constant along a magnetic flux tube provided that the system is in a steady state. For our spherically symmetric models this reduces to:
\begin{equation}\label{eq:streamingpolytropic}
    P_\mathrm{c} = P_\mathrm{c0} \lr{\frac{\rho}{\rho_\mathrm{0}}}^{\gamma/2}
\end{equation}
where $\gamma/2 = 2/3$ and we set $P_\mathrm{c0} = P_\mathrm{c}(R)$ and $\rho_\mathrm{0} = \rho(R)$.

In the extrinsic turbulence model \citep{zweibelreview2017}, the cosmic rays are scattered by waves that are a result of a turbulent cascade or other physical process. In this model, the cosmic rays still transfer momentum through their pressure gradient but no energy transfer occurs and $\mbfvA$ disappears from eqn.(\ref{eq:fulltransport}). This model can accurately model diffusion and so assuming $\mathbf{u} = 0$ and $Q$ is negligible, our transport equation for pure diffusion is:
\begin{equation}
    - \bnabla\cdot\lr{\kappa \bnabla P_\mathrm{c}} =  0
\label{eq:purediff}
\end{equation}
If we move to a spherical coordinate system and assume that $\kappa$ is uniform throughout the galaxy, we can solve eqn. (\ref{eq:purediff}) and find the cosmic ray pressure goes as:
\begin{equation}
    P_\mathrm{c}(r) = P_\mathrm{c0} \lr{\frac{R}{r}}
\end{equation}
assuming that $P_\mathrm{c}(R) = P_\mathrm{c0}$ and $P_\mathrm{c}'(R) = P_\mathrm{c0}/R$ where $R$ is the outer radius of our central mass distribution. 

The extrinsic turbulence model can also lead to advection in the limit of small diffusivity. In this model, any motion of the cosmic rays must only be due to the motion of the surrounding medium. In the pure advection case, therefore, we drop our diffusion term and ignore the $Q$ term to derive:
\begin{equation}
    \bnabla\cdot(\gamma \mathbf{u} P_\mathrm{c}) = (\gamma - 1)\mathbf{u}\cdot\bnabla P_\mathrm{c}
\label{eq:generalAdvect}
\end{equation}

Similar to the streaming case, we can simplify eqn. (\ref{eq:generalAdvect}) to get a relation between the cosmic ray pressure and the gas density. In a steady state, $\rho\bnabla\cdot \mathbf{u} = -\mathbf{u}\cdot \bnabla \rho$ so if we substitute this relation into eqn. (\ref{eq:generalAdvect}) and drop sources and collisions, we can solve for $P_\mathrm{c}$ and find:
\begin{equation}\label{eq:advectionpolytropic}
    P_\mathrm{c} = P_\mathrm{c0} \lr{\frac{\rho}{\rho_\mathrm{0}}}^\gamma
\end{equation}
which we note is then independent of the velocity of the fluid. Therefore, we can see that the just advection case follows a polytropic relation where the cosmic rays behave just like a relativistic gas.

\subsection{Sources and Losses}
\label{subsec:setupSC}

Finally, in \S\ref{subsec:numResults}, we will add sources 
and losses to our analysis. These effects are represented by the term $Q$ in eqn. (\ref{eq:fulltransport}) and account for the effects of supernovae  (the sources) and hadronic collisions (the losses). As explained below, in the region $r < R$, we also include a loss term due to diffusive transport. We write $Q$ in the general form:
\begin{equation}
Q = S - \frac{U_\mathrm{c}}{\tau_\mathrm{L}}
\label{eq:definingQ}
\end{equation}
where $S$ is the cosmic ray source density, $U_\mathrm{c}$ is the cosmic ray energy density, and $\tau_\mathrm{L}$ is the loss time

\subsubsection{Source Term}
To implement sources, we assume that the cosmic rays are injected into our system by supernovae which occur at a rate that follows the Kennicutt-Schmidt (K-S) Law for star formation \citep{KennicuttSFR1998}. Equation (4) of \cite{KennicuttSFR1998} gives the star formation rate per area $\Sigma_\mathrm{SFR}$ (in $\unit M_{\odot} \unit kpc^{-2} \unit yr^{-1}$) as a function of gas surface density $\Sigma_\mathrm{gas}$ (in $ \unit M_{\odot} \unit pc^{-2}$); adopting mean values this is:
\begin{equation}
\Sigma_\mathrm{SFR}=2.5\times 10^{-4}\Sigma_\mathrm{gas}^{1.4}.
\label{eq:ksLaw}
\end{equation}
That is, the K-S law is of the form $\Sigma_\mathrm{SFR}=K\Sigma_\mathrm{gas}^\alpha$. It will be important to note that the exponent used in the K-S Law, $\alpha$, has varied since the foundational paper \citep{Schmidt59} which proposed that the SFR is proportional to $\rho^2$ (i.e to volume density, not, surface density)  and then over the past two decades since \cite{KennicuttSFR1998}. For example, \cite{LiuKSRevision2015} found that $\alpha$ can be $1.01$, $1.12$, or $1.62$ depending on the assumption of the rate of conversion from CO to $\rm H_2$, while \cite{KennicuttSFRLaw2021} updated the original K-S Law with an $\alpha = 1.5$. Therefore, $\alpha$ will be a varied parameter in our analysis in \S\ref{subsec:numResults}. The implementation of the K-S Law into our numerical equations is further explained in Appendix \ref{sec:nondimens}.

We derive an equation for $S$ in terms of gas density in the starburst region by assuming star formation takes place in a layer of thickness $z$,
 that the mass in  stars required to produce one supernova is $m_\mathrm{SN}$, and that each supernova
 produces energy $\epsilon_c$ in cosmic rays.
Generally, we will assume $z=200$ pc,  $m_\mathrm{SN} = 100 \unit M_{\odot} \unit SN^{-1}$ \citep{MannucciSFRMassRate2005}, and $\epsilon_c=10^{50}$ erg. Note that we have chosen a fairly optimistic value for the supernova rate per unit mass from \cite{MannucciSFRMassRate2005}, although it can vary greatly based on the age, luminosity, and galaxy type. With these assumptions, the source function
$S$ can be written in terms of gas density as
\begin{equation}
S = (1.9\times10^{-72})z^{1.4}\rho_\mathrm{0}^{1.4} \epsilon_\mathrm{c} \unit ergs \unit cm^{-3} \unit s^{-1}
\label{eq:sourceterm}
\end{equation}

\subsubsection{Loss Terms}
In the region $r > R$, where we solve for the structure of the atmosphere and adopt an explicit transport model, we assume losses are
due entirely to hadronic collsions, which occur at rate $\tau_C$. From \cite{CrockerEddington2021Part1}, the collision time can be written as:
\begin{equation}
    \tau_\mathrm{C} = 100 {\unit Myr} \lr{\frac{10^{-24}}{\rho}} = \frac{3.2\times 10^{-9}}{\rho} \unit s.
\label{eq:KSLawCollisionTime}
\end{equation}

In $r < R$, we also include a diffusive transport term $\tau_T$, where we envision diffusion as due to propagation on tangled fieldlines along which cosmic rays are scattered by small scale, small amplitude fluctuations. 
We take for the transport time:
\begin{equation}
    \tau_{T} = \frac{z^2}{\kappa}
\end{equation}
where $\kappa$ is the diffusivity. 
The loss rates due to collisions and transport are additive, and lead to a loss time $\tau_L$: 
\begin{equation}
\tau_L=\frac{\tau_C\tau_T}{\tau_C+\tau_T}.
\end{equation}
Using eqn. (\ref{eq:KSLawCollisionTime}), we have:
\begin{equation}
    \tau_\mathrm{L} = 3.2\times10^{-9} \frac{z^2}{(3.2\times10^{-9}\kappa + \rho_\mathrm{0} z^2)}
\label{eq:CRlifetime}
\end{equation}


\subsubsection{Cosmic ray pressure in the core}
We determine conditions in the core by assuming sources balance losses. From eqns. (\ref{eq:sourceterm}) and (\ref{eq:CRlifetime}):
\begin{equation}
U_\mathrm{c0}= (6.08\times10^{-81})z^{3.4} \frac{\rho_\mathrm{0}^{1.4} \epsilon_\mathrm{c}}{(3.2\times10^{-9}\kappa + \rho_\mathrm{0} z^2)}
\label{eq:Uc0fromKS}
\end{equation}
in erg cm$^{-3}$ It is straightforward to adapt eqn. (\ref{eq:Uc0fromKS}) to any exponent $\alpha$ in the star formation law. The prefactor, of course, will change, but the factor of $z^{3.4}$ should be read as $z^{\alpha + 2}$, and the factor of $\rho_0^{1.4}$ should be read as $\rho_0^{\alpha}$.

We can check the reliability of our formulae by applying it to the starburst core of M82, for which many of the relevant parameters are observable or constrained by independent modeling. Applying eqn. (\ref{eq:ksLaw}) for the parameters given
in Table 1, we predict an SFR of 4.9 $M_{\odot}$/yr, which is lower by a factor of 2 than the accepted value of 10$M_{\odot}$/yr. Using the Milky Way value, $\kappa = 3\times10^{28} \unit cm^2 \unit s^{-1}$ \citep{StrongCRProp2007} in eqn. (\ref{eq:Uc0fromKS}) and again taking $z=200$ pc gives $U_{c0}=210$ eV cm$^{-3}$, also about a factor of 2 less than the best fit value of 525 eV cm$^{-3}$
obtained in \cite{yoasthullm82_2013}, agreement we regard as satisfactory given the many assumptions 
underlying the analysis and models.

It may be more appropriate to  derive $U_\mathrm{c0}$ directly from the star formation rate if the latter is known. In that case, we can define the cosmic ray luminosity due to star formation as:
\begin{equation}
    L_\mathrm{SFR} = \frac{SFR \epsilon_\mathrm{c}}{m_\mathrm{SN}}
\end{equation}
where $SFR$ is the star formation rate in $\rm M_{\odot} \unit yr^{-1}$. The source density $S$ then is just $L_\mathrm{c}/V_{enc}$ where we will assume that again the star formation is occurring in a disk close to the midplane of the galaxy. Therefore:
\begin{equation}
    S = \frac{SFR \epsilon_\mathrm{c}}{m_\mathrm{SN} \pi R^2 z}
\end{equation}
Finally, to obtain $U_\mathrm{c0}$, we take $S \tau_\mathrm{L}$ where $\tau_\mathrm{L}$ is the same as defined in eqn. (\ref{eq:CRlifetime}):
\begin{equation}
    U_\mathrm{c0} = 1.01\times10^{-16} \frac{SFR \epsilon_\mathrm{c} z}{m_\mathrm{SN} \pi R^2 (3.2\times10^{-9}\kappa + \rho_\mathrm{0} z^2)} \unit ergs \unit cm^{-3}
\label{eq:Uc0fromSFR}
\end{equation}
Since the SFR is known for all the galaxies analyzed in this paper, we will use eqn. (\ref{eq:Uc0fromSFR}) and will use it to move between $U_\mathrm{c0}$ and $SFR$ when needed.

We have already seen two transport models which lead to polytropic relationships between $P_c$ and $\rho$ (eqns. \ref{eq:streamingpolytropic} and \ref{eq:advectionpolytropic}). Yet another polytropic relation holds if sources and collisional losses dominate, which we term the calorimetric limit. From eqn. (\ref{eq:definingQ})
\begin{equation}
    S(\rho) = \frac{U_\mathrm{c}}{\tau_\mathrm{C}}
\end{equation}
in this case. We can use the the Kennicutt-Schmidt Law ($\Sigma_\mathrm{SFR} = K \Sigma_\mathrm{gas}^{\alpha}$) to obtain the form of our source term while we can calculate the value of $\tau_\mathrm{C}$ from eqn. (\ref{eq:KSLawCollisionTime}). Thus, we have:
\begin{equation}
    S_\mathrm{0}\lr{\frac{\rho}{\rho_\mathrm{0}}}^{\alpha} = \frac{3P_\mathrm{c} \rho}{\tau_\mathrm{C,0}\rho_\mathrm{0}}
\label{eq:CalLimitDeriv}
\end{equation}
where $\tau_\mathrm{C,0} = 3.2\times10^{-9}$. Solving eqn. (\ref{eq:CalLimitDeriv}) for $P_\mathrm{c}$, we find:
\begin{equation}
    P_\mathrm{c} = \frac{S_\mathrm{0} \tau_\mathrm{C,0}}{3}\lr{\frac{\rho}{\rho_\mathrm{0}}}^{\alpha - 1} = P_\mathrm{c0}\lr{\frac{\rho}{\rho_\mathrm{0}}}^{\alpha - 1}
    \label{eq:calorimetricPolytrop}
\end{equation}

\subsection{Cosmic ray supported atmospheres with polytropic equations of state}
\label{subsec:transportModels}

In order to show the methodology of our work a little more clearly, we will first solve for the Eddington limit of a more simple system where a numerical solver will not be required. 

As shown in \S\S \ref{subsec:CRTransport} and \ref{subsec:setupSC}, the streaming, advection, and calorimetric limit models will obey a polytropic relationship where:
\begin{equation}
    P_\mathrm{c} = P_\mathrm{c0}\lr{\frac{\rho}{\rho_\mathrm{0}}}^{a}
    \label{eq:polytrop}
\end{equation}
where $a$ is a general exponent that is related to the familiar polytropic index $n$ from stellar structure theory by $n=(a-1)^{-1}$. 

We will treat the core quantities $P_\mathrm{c0}$ and $\rho_\mathrm{0}$ as our starting (base) values for the cosmic ray pressure and gas density respectively (where $P_\mathrm{c0} = U_\mathrm{c0}/3$). We can then substitute eqn. (\ref{eq:polytrop}) into eqn. (\ref{eq:hydroequil}) for a general potential $\Phi$:
\begin{equation}
    \frac{d}{dr}\lr{P_\mathrm{c0}\lr{\frac{\rho}{\rho_\mathrm{0}}}^{a}} = -\rho\lr{\frac{d\Phi}{dr}} 
\label{eq:unsimpleGeneral}
\end{equation}
from which $\rho(r)$ is found to be
\begin{equation}
    \rho(r) = \rho_\mathrm{0}\lr{1 + \frac{a - 1}{a}\frac{\rho_\mathrm{0}}{P_\mathrm{c0}}\lr{\Phi(R) - \Phi(r)}}^{1/(a - 1)}
    \label{eq:general_density}
\end{equation}
Equation (\ref{eq:general_density}) is valid for any spherically symmetric potential $\Phi$, but when evaluating it we will always assume $\Phi$ is given by eqn. (\ref{eq:halomasspot}) so that $\Phi(R) = -GM_c/R - GM_h/(R+a)$.

Substituting eqn. (\ref{eq:general_density}) into the polytropic relation from eqn. (\ref{eq:polytrop}), we have:
\begin{equation}
    P_\mathrm{c}(r) = P_\mathrm{c0}\lr{1 + \frac{a - 1}{a}\frac{\rho_\mathrm{0}}{P_\mathrm{c0}}\lr{\Phi(R) - \Phi(r)}}^{a/(a - 1)}
\label{eq:generalPc}
\end{equation}

For eqn. (\ref{eq:general_density}), we can see that assuming $\Phi\rightarrow 0$ as $r\rightarrow\infty$, the density at infinity satisfies:
\begin{equation}
    \mathbf{\rho_\mathrm{\infty}=\rho_{0}\lr{1+\frac{a - 1}{a}\frac{\rho_\mathrm{0}\Phi(R)}{P_\mathrm{c0}}}^{1/(a - 1)}}
\label{eq:generalAsymptoteRho}
\end{equation}
while similarly for eqn. (\ref{eq:generalPc}):
\begin{equation}
P_\mathrm{c\infty}=P_\mathrm{c0}\left[1+\frac{a - 1}{a}\frac{\rho_\mathrm{0}\Phi(R)}{P_\mathrm{c0}} \right]^{a/(a - 1)}.
\label{eq:generalAsymptotePc}
\end{equation}
Note that the quantity $\rho_\mathrm{0} \Phi(R) / P_\mathrm{c0}$ will appear in many of our equations throughout this paper. It is a quantity that essentially represents the balance of cosmic ray pressure against the force of gravity. If we note that the squared cosmic ray sound speed, $v_\mathrm{c0}^2 \propto P_\mathrm{c0} / \rho_\mathrm{0}$, and squared escape velocity, $v_\mathrm{esc}^2 
\propto \vert\Phi(R)\vert$, then we can rewrite this quantity as:
\begin{equation}
    \frac{\rho_\mathrm{0}\vert\Phi(R)\vert}{P_\mathrm{c0}} \propto \left(\frac{v_\mathrm{esc}}{v_\mathrm{c0}}\right)^2.
\end{equation}
Therefore, $\rho_0\Phi(R)/P_{c0}$ is  a good measure of how well confined a system is and its value will be of particular interest for us in understanding the galaxies we study. It appears in the dimensionless equations introduced in Appendix \ref{sec:nondimens} as the parameter $\epsilon$.

Following Parker's argument for the necessity of the solar wind \citep{ParkerSolarWind1958a}, we will find the asymptotic value of the cosmic-ray pressure for the extrinsic turbulence, self-confinement, and calorimetric models. If this asymptotic value occurs above the assumed pressure of the CGM, then we can definitively say that system is super-Eddington. For a typical value of the CGM pressure, we have referred to \cite{JiCGMwithCRs2020} and find an approximate average value of $P_\mathrm{CGM} \sim 10^{-15} \unit ergs \unit cm^{-3}$ for the thermal pressure. 

We can conclude the Eddington cosmic ray pressure will be the initial pressure that results in the pressure asymptoting at $P_\mathrm{CGM}$. Defining this as $P_\mathrm{Edd}$ and setting the asymptotic pressure to $P_\mathrm{CGM}$ in eqn. (\ref{eq:generalAsymptotePc}):
\begin{equation}
        P_\mathrm{CGM}=P_\mathrm{Edd}\lr{1 + \frac{a - 1}{a}\frac{\rho_\mathrm{0}\Phi(R)}{P_\mathrm{Edd}}}^{a/(a-1)}
\label{eq:PcEdd}
\end{equation}

\subsubsection{Advection}
\label{subsec:crturbAna}
For advection, we showed in \S\ref{subsec:CRTransport} that $a = \gamma = 4/3$. Substituting $a = 4/3$ into eqn. (\ref{eq:generalAsymptoteRho}):
\begin{equation}
    \rho_\mathrm{\infty} = \rho_\mathrm{0}\lr{1 + \frac{\rho_\mathrm{0}\Phi(R)}{4P_\mathrm{c0}}}^{3}
\end{equation}
and eqn. (\ref{eq:generalAsymptotePc}):
\begin{equation}
    P_\mathrm{c\infty} = P_\mathrm{c0}\lr{1 + \frac{\rho_\mathrm{0}\Phi(R)}{4P_\mathrm{c0}}}^{4}
\end{equation}
We can see here that $P_\mathrm{c\infty}< 0$ if $P_\mathrm{c0} < \rho_\mathrm{0}\vert\Phi(R)\vert/4$, which we can interpret as the gas always being confined.

To obtain the cosmic ray pressure needed for the Eddington limit, we use eqn. (\ref{eq:PcEdd})
\begin{equation}
    P_\mathrm{CGM} = P_\mathrm{Edd}\lr{1 + \frac{\rho_\mathrm{0} \Phi(R)}{4P_\mathrm{Edd}}}^{4}
\label{eq:PcadvEdd}
\end{equation}
Equation (\ref{eq:PcadvEdd}) is best solved numerically for $P_\mathrm{Edd}$ but we can make an analytical approximation to the solution by noting that generally, we expect the asymptotic pressure, $P_\mathrm{CGM}$ to be much smaller than the base pressure at the edge of our mass distribution. Assuming $P_\mathrm{CGM} \ll P_\mathrm{c0}$, we find
\begin{equation}\label{eq:PcadvEddexpression}
P_\mathrm{Edd}\sim\frac{\rho_\mathrm{0}\vert \Phi(R)\vert}{4}.
\end{equation}
. 

\subsubsection{Cosmic Ray Streaming}
\label{subsec:crstreamAna}
In the self-confinement model, we know from eqn. (\ref{eq:streamingpolytropic}) that $a = 2/3$ and thus:
\begin{equation}
     \rho_\mathrm{\infty} = \rho_\mathrm{0}\lr{1 - \frac{\rho_\mathrm{0} \Phi(R)}{2P_\mathrm{c0}}}^{-3}
\end{equation}
for the density and:
\begin{equation}
    P_\mathrm{c\infty} = P_\mathrm{c0}\lr{1 - \frac{\rho_\mathrm{0}\Phi(R)}{2P_\mathrm{c0}}}^{-2}
\label{eq:streamingPc}
\end{equation}
for the cosmic ray pressure. For this case, contrary to the advection case, $P_\mathrm{c\infty}$ is always positive, so the
radius of the cosmic ray supported envelope is limited only by the confining pressure of the CGM.

An approximate solution of eqn. (\ref{eq:PcEdd}), valid for $\rho_0\vert\Phi(R)\vert/P_\mathrm{CGM}\gg 1$, is
\begin{equation}
    P_\mathrm{Edd} = P_\mathrm{CGM}^{1/3}\lr{\frac{\rho_\mathrm{0}\vert\Phi(R)\vert}{2}}^{2/3}
\label{eq:streamingPEdd}
\end{equation}

\begin{table*}[t!]
    \centering
    \begin{tabular}{|c||c|c|c|c|c|c|c|c|c|c|}
    \hline
       Galaxy & $M_\mathrm{c}$ & $M_{halo}$ & $R_{dyn}$ & $a$ & $\rho_{0}^{gas}$ & $\rho_{0}^{ion} $ & $B_\mathrm{0}$ & $SFR$ & $U_\mathrm{C,0}$ & $\rho_\mathrm{0} \vert\Phi(R)\vert /P_\mathrm{c0}$\\
       \hline
       MW & $1.4 \times 10^{9}$\tablenotemark{a} & $1.3 \times 10^{12}$\tablenotemark{b} & $0.23$\tablenotemark{c} & $16.6$ & $1249$\tablenotemark{c} & $16.7$\tablenotemark{c} & $10$\tablenotemark{d} & 0.01\tablenotemark{e} & $10$\tablenotemark{f} & $8.56\times10^{5}$ \\
       \hline
       M82 & $1\times10^{9}$\tablenotemark{g,h} & $5.5\times10^{11}$\tablenotemark{i} & $0.2$\tablenotemark{h} & $13.0$ & $1580$\tablenotemark{h} & $167$\tablenotemark{h} & $300$\tablenotemark{j} & $10$\tablenotemark{h} & $525$\tablenotemark{j} & $1.26\times10^{4}$\\
       \hline
       LMC & $4\times10^{9}$\tablenotemark{k} & $2\times10^{11}$\tablenotemark{$\ell$} & $1.7$\tablenotemark{k} & $8.84$ & $7.9$\tablenotemark{k} & $3.34$\tablenotemark{k} & $4$\tablenotemark{k} & $0.4$\tablenotemark{k} & $0.58$\tablenotemark{j} & $3.35\times10^{4}$ \\
       \hline
       NGC 4449 & $2.1\times 10^{9}$\tablenotemark{m} & $2\times10^{11}$\tablenotemark{m} & $1.83$\tablenotemark{m} & $8.84$ & $4.90$\tablenotemark{n} & $3.40$\tablenotemark{n} & $12$\tablenotemark{o} & $0.97$\tablenotemark{m} & $3.82$ & $3.32\times10^{3}$\\
       \hline
       DRC-8 & $8.2\times10^{10}$\tablenotemark{p} & $9\times10^{12}$\tablenotemark{p} & $10$\tablenotemark{p} & $24.6$\tablenotemark{q} & $3950$\tablenotemark{q} & $52.8$\tablenotemark{q} & $1000$\tablenotemark{q} & $394$\tablenotemark{p} & $3.18$ & $1.3\times10^{8}$ \\
       \hline
    \end{tabular}
    \caption{The various dynamical masses ($M_{\odot}$), halo masses ($M_{\odot}$), radii confining the dynamical mass (kpc), the scale height for the dark matter halo (kpc), gas and ion mass densities (in units of $10^{-24} \rm g \unit cm^{-3}$), base magnetic fields ($\mu G$), star formation rates ($\rm M_\odot \unit yr^{-1}$), and cosmic ray energy densities ($\rm eV \unit cm^{-3}$) for each galaxy that we model. The last column includes the value of $\rho_\mathrm{0} \Phi(R)/ P_\mathrm{c0}$ for each galaxy which as mentioned in \S\ref{subsec:transportModels} will be a particular quantity of interest for this work. For $\Phi(R)$, we have assumed our potential is the Hernquist potential, shown in eqn. (\ref{eq:halomasspot}), where $\Phi(R) = \Phi(R_{dyn})$ in that equation. Note that $P_\mathrm{c0} = U_\mathrm{c0}/3$. For the Alfv\'{e}n speed, we need to find the ion density specifically as the waves which scatter the cosmic rays are sufficiently high frequency that the plasma and neutrals are decoupled. For some galaxies, like our Galaxy and M82, we have tried to constrain ourselves to their central molecular zones (CMZs) while for the other three galaxies, we have generally incorporated the entire galaxy into our potential well. Our values for $R_{dyn}$ reflect these differences in the potential wells}. Finally, we have obtained the gas and ion mass densities from the gas and ion surface densities provided in Table 2 of \cite{McQuinnDwarfStarburst2012}.
    \label{tab:galacticparams}

    \tablenotemark{a}{\cite{LaunhardtNBMassofMW2002}},
    \tablenotemark{b}{\cite{PostiDMHaloMass2019}},
    \tablenotemark{c}{\cite{FerriereISMGas2007}},
    \tablenotemark{d}{\cite{GuenduezCMZMagnetic2020}},
    \tablenotemark{e}{\cite{yoasthullMWCMZ2014}},
    \tablenotemark{f}{\cite{everettwinds2008}},
    \tablenotemark{g}{\cite{MayyaM82Summary2009}},
    \tablenotemark{h}{\cite{yoasthullm82_2013}},
    \tablenotemark{i}{\cite{OehmM81GroupEvol2017}},
    \tablenotemark{j}{\cite{yoasthullstarbursts2016}},
    \tablenotemark{k}{\cite{BustardLMCWinds2020}},
    \tablenotemark{$\ell$}{\cite{LucchiniLMCStream2020}}
    \tablenotemark{m}{\cite{McQuinnDwarfStarburst2019}},
    \tablenotemark{n}{\cite{McQuinnDwarfStarburst2012}},
    \tablenotemark{o}{\cite{ChyzyNGC4449_2000}},
    \tablenotemark{p}{\cite{LongYoungMassiveGalaxies2020}},
    \tablenotemark{q}{Arianna Long personal communication}
\end{table*}

\subsubsection{Calorimetric Limit}
In the calorimetric case, we will use two different K-S Laws, one where $\alpha = 2$ and another where $\alpha = 1.4$.  The corresponding polytropic exponent will be $a = 1$ and $a = 0.4$ respectively.

In the first case, $\alpha = 2$, we get an isothermal polytropic relation where $P_\mathrm{c} = P_\mathrm{c0} (\rho/\rho_\mathrm{0})$ and $P_{c0}=S_0\tau_{c0}/3$. We cannot use our general asymptotic form for $\rho$ and $P_\mathrm{c}$ from eqns. (\ref{eq:generalAsymptoteRho}) and (\ref{eq:generalAsymptotePc}) here and instead we find that the density is:
\begin{equation}
    \rho(r) = \rho_\mathrm{0} e^{-\rho_\mathrm{0}(\Phi(r) - \Phi(R))/P_\mathrm{c0}}
\end{equation}
with a cosmic ray pressure of:
\begin{equation}
    P_\mathrm{c}(r) = P_\mathrm{c0} e^{-\rho_\mathrm{0}(\Phi(r) - \Phi(R))/P_\mathrm{c0}}.
\end{equation}
The asymptotic values of $\rho$ and $P_c$ are:
\begin{equation}
    \rho_\mathrm{\infty} = \rho_\mathrm{0} e^{\rho_\mathrm{0}\Phi(R)/P_\mathrm{c0}},
\end{equation}
\begin{equation}
    P_{c,\infty} = P_\mathrm{c0} e^{\rho_\mathrm{0}\Phi(R)/P_\mathrm{c0}}.
\end{equation}
Setting $P_{c\infty}=P_\mathrm{CGM}$, $P_{c0}=P_\mathrm{Edd}$ leads to the transcendental equation for $P_\mathrm{Edd}$
\begin{equation}
 \frac{P_\mathrm{Edd}}{P_\mathrm{CGM}} =e^{-{\rho_0\Phi(R)/P_\mathrm{Edd}}}
\label{eq:PEDDcal2}
\end{equation}
In the other case, $\alpha = 1.4$, we find:
\begin{equation}
        \rho_\mathrm{\infty}=\rho_{0}\lr{1-\frac{3}{2}\frac{\rho_\mathrm{0}\Phi(R)}{P_\mathrm{c0}}}^{-3/5}
\end{equation}
and:
\begin{equation}
    P_\mathrm{c\infty}=P_\mathrm{c0}\left[1-\frac{3}{2}\frac{\rho_\mathrm{0}\Phi(R)}{P_\mathrm{c0}} \right]^{-2/3}
    \label{eq:asympt_pres_cal}
\end{equation}
Assuming that $P_\mathrm{c\infty} = P_\mathrm{CGM}$ and that $P_\mathrm{c0} = P_\mathrm{Edd}$, eqn. (\ref{eq:asympt_pres_cal}) becomes:
\begin{equation}\label{eq:PEDDcal1.4}
        P_\mathrm{CGM}=P_\mathrm{Edd}\left[1-\frac{3}{2}\frac{\rho_\mathrm{0}\Phi(R)}{P_\mathrm{Edd}} \right]^{-2/3}
\end{equation}
which has the approximate solution:
\begin{equation}
P_\mathrm{Edd}\approx P_\mathrm{CGM}^{3/5}\left(\frac{3}{2}\rho_0\vert\Phi(R)\vert\right)^{2/5} 
\label{eq:PEDDcal1.4approx}
\end{equation}

\subsubsection{Diffusion}
\label{subsec:crdiffAna}
In the case of diffusive transport, a polytropic relationship cannot be defined. Instead, we have that:
\begin{equation}
    \frac{1}{r^2}\frac{d}{dr}\lr{\kappa r^2 \frac{dP_\mathrm{c}}{dr}} = 0
\end{equation}
for a spherical system.  If we assume that $\kappa$ is uniform throughout the galaxy, the cosmic ray pressure goes as:
\begin{equation}
    P_\mathrm{c}(r) = P_\mathrm{c0} \lr{\frac{R}{r}}
\label{eq:diffusion_pressure}
\end{equation}
assuming that $P_\mathrm{c}(R) = P_\mathrm{c0}$ and $P_\mathrm{c}'(R) = -P_\mathrm{c0}/R$. Substituting into eqn. (\ref{eq:hydroequil}), we get a density of:
\begin{equation}
    \rho(r) = P_{c0}\lr{\frac{d\Phi}{dr}}^{-1} = \frac{P_\mathrm{c0}}{\Phi_\mathrm{0}}\lr{1+\frac{\mu}{(1+a/r)^2}}^{-1}
\label{eq:diffusion_density}
\end{equation}
where we define $\Phi_0 = GM_c/R$. We note that this density would be constant without a dark matter halo. It is also important to note then that the base density for just diffusion is:
\begin{equation}
    \rho_\mathrm{0} = \frac{P_\mathrm{c0}}{\Phi_\mathrm{0}}\lr{1+\frac{\mu}{(1+a/R)^2}}^{-1}
    \label{eq:anaDiffrho0}
\end{equation}

We can first note from eqn. (\ref{eq:diffusion_pressure}) that the pressure will always approach zero as $r \rightarrow \infty$. Therefore, our definition of the asymptotic value being larger than $P_\mathrm{CGM}$ will never work here. Thus, in this case, we could define a CGM radius, say $R_\mathrm{CGM}$ at which if the cosmic ray pressure is still above the CGM pressure the system could be considered super-Eddington. Therefore, we find that $P_\mathrm{Edd}$ is just given by:
\begin{equation}
    P_\mathrm{Edd} = P_\mathrm{CGM}\frac{R_\mathrm{CGM}}{R}
\label{eq:anaDiffeddPres}
\end{equation}

To further illuminate this model, let's run a quick check of its results for M82, a starburst galaxy, whose parameters can be found in Table \ref{tab:galacticparams}. First, let's assume that instead of going to infinity, the boundary between the ISM and the CGM occurs at $20 \unit kpc$. If we just solve for the Eddington cosmic ray pressure, we find that $P_\mathrm{Edd} = 10^{-13} \unit dynes \unit cm^{-2}$, equivalent to a cosmic ray energy density of $U_\mathrm{Edd} = 0.19 \unit eV \unit cm^{-3}$, about three orders of magnitude lower than M82's actual $U_\mathrm{c0}$. Therefore, at a first glance, diffusion seems like a great candidate to reach a cosmic ray Eddington limit. However, if we then solve for $\rho_\mathrm{0}$ in eqn. (\ref{eq:anaDiffrho0}) using this $P_\mathrm{Edd}$ as our $P_\mathrm{c0}$, we find that the corresponding value is $\rho_\mathrm{Edd} = 4.11\times10^{-28} \unit g \unit cm^{-3}$. This is extremely small and obviously not physically reasonable for any galaxy, including M82. 

It's worth stating that while the assumption of constant $\kappa$ is surely unrealistic, allowing $\kappa$ to scale as a power of density, as assumed in \cite{CrockerEddington2021Part1}, is more general but equally \textit{ad hoc}, and produces a density which drops off extremely slowly with $r$.

To illustrate further that a variable diffusion coefficient also creates problems for reaching an Eddington limit, we now assume that $\kappa = \kappa_0 (r/R)^\beta$ where $\beta$ is some constant that can be chosen freely (we assume $\beta\ge 0$). Plugging this form for $\kappa$ into eqn. (\ref{eq:purediff}), we find:
\begin{equation}
    r^2 \kappa \frac{d P_c}{dr} = const.
\end{equation}
which when integrated gives:
\begin{equation}
    P_c = P_{c0} \lr{\frac{r}{R}}^{-(1+\beta)}
    \label{eq:variablediffusion_pressure}
\end{equation}
where we have assumed that $P_c(R) = P_{c0}$ and used $P_c'(R) = -P_{c0}(1+\beta)\lr{\frac{r}{R}}^{-(2+\beta)}$. Proceeding as we
did to derive eqn. (\ref{eq:anaDiffeddPres}) we find an Eddington pressure of:
\begin{equation}
    P_\mathrm{Edd} = P_\mathrm{CGM}\lr{\frac{R_\mathrm{CGM}}{R}}^{(1+\beta)}
    \label{eq:variablediffusion_Eddpressure}
\end{equation}
We can see that for $\beta = 0$, we arrive back at the solution for our constant diffusion coefficient.

We can then define the Eddington densities for this model by assuming that our base pressure is the Eddington pressure, $P_\mathrm{Edd}$. Therefore, we can plug eqn. (\ref{eq:variablediffusion_Eddpressure}) into eqn. (\ref{eq:hydroequil}) and solving for $\rho(r)$:
\begin{equation}
    \rho(r) = \frac{(1+\beta)P_\mathrm{CGM}}{R_\mathrm{CGM}}\lr{\frac{d\Phi}{dr}}^{-1}\lr{\frac{R_\mathrm{CGM}}{r}}^{(2+\beta)}
\end{equation}
which gives a base density of:
\begin{equation}
    \rho_\mathrm{Edd} = \frac{(1+\beta)P_\mathrm{CGM}}{R_\mathrm{CGM}}\lr{\frac{d\Phi}{dr}\bigg\vert_R}^{-1}\lr{\frac{R_\mathrm{CGM}}{R}}^{(2+\beta)}
    \label{eq:variablediffusion_Edddensity}
\end{equation}

If we write eqn. (\ref{eq:variablediffusion_Edddensity}) in terms of our density for constant diffusion (eqn. (\ref{eq:anaDiffrho0}), we derive:
\begin{equation}
    \rho_\mathrm{Edd}(\beta) = \rho_\mathrm{Edd}(0) (1+\beta)\lr{\frac{R_\mathrm{CGM}}{R}}^{\beta}
    \label{eq:variablediffusion_densitiesequation}
\end{equation}
We recall that for M82 with constant diffusion, we found that $\rho_\mathrm{Edd} = 4.11\times 10^{-28} \unit g \unit cm^{-3}$. Taking the parameters of M82 and plugging those into eqn. (\ref{eq:variablediffusion_densitiesequation}) to find an Eddington density equivalent to M82's density from Table \ref{tab:galacticparams}, we solve and find that $\beta = 3.012$ gives us the exact density for M82. However, we note that for this beta, we can solve eqn. (\ref{eq:variablediffusion_Eddpressure}) and find that $P_\mathrm{Edd} = 1.057\times10^{-7} \unit dynes \unit cm^{-2}$. This Eddington pressure is equivalent to a base cosmic ray energy density of $U_\mathrm{Edd} = 1.98\times10^5 \unit eV \unit cm^{-3}$ which is well above the actual cosmic ray energy density for M82. We can see from this model then that if we get a more reasonable base density, we get a huge Eddington cosmic ray energy density that the galaxy would never be able to obtain. Therefore, regardless of the model we use for diffusion, we find in general that it does not produce a system capable of reaching the Eddington limit.

Due to these results and to our finding that sources and collisions do not heavily affect these conclusions (to be shown in \S\ref{subsec:numResults}), we will ignore diffusion as a method of cosmic ray transport for the rest of this paper and instead largely focus on streaming.

We summarize the results of these transport models as follows. The quantity $\rho_0\Phi(R)/P_{c0}$ appears in all cases as a figure of merit that represents the depth of the gravitational potential well relative to the energy per mass available for driving an outflow. At first glance, we might expect that this parameter must be of order unity or less in a super-Eddington galaxy. For the advective and $\alpha = 2$ calorimetric cases, this is indeed true. But as eqns. (\ref{eq:streamingPEdd}) and (\ref{eq:PEDDcal1.4approx}) show, under certain conditions - transport by streaming or calorimetry with an $\alpha = 1.4$ star formation law in the cases studied here - substantially smaller base cosmic ray pressures can unbind the system as well. In terms of the polytropic index $n$, systems with $n > 0$ have $P_\mathrm{Edd}\sim\rho_0\vert\Phi(R)\vert$ while systems with $n < 0$ have $P_\mathrm{Edd}\sim P_\mathrm{CGM}^{-1/n}(\rho_0\vert\Phi(R)\vert)^{1+1/n}$. 

It will now be useful to apply these models to galaxies for which the input parameters are known. 

\subsection{Parameter Values}
\label{subsec:parameters}
For this work, we will want to examine the viability of a wind being launched solely by cosmic rays for a few different galaxy types. Therefore, we have chosen five different galactic models to analyze for the rest of this work, represented in Table \ref{tab:galacticparams}.

The five galaxies we have chosen are the Milky Way (a large, older, spiral galaxy), M82 (a starburst galaxy), the Large Magellanic Cloud (LMC) (a dwarf galaxy), NGC 4449 (a dwarf starburst galaxy), and DRC-8 (a massive, young, starbursting galaxy). It is important to note that winds have been observed for M82 \citep{OconnellM82Central1978, StricklandM82Wind2009} and NGC 4449 \citep{McQuinnDwarfStarburst2019}. There is also evidence for outflows from the LMC \citep{StaveleySmithLMCHI2003, BargerLMCWind2016} and some analyses have interpreted the observation of soft, diffuse X-ray emission in some regions of the Milky Way as evidence for a wind \citep{everettwinds2008}.
For the penultimate column in Table \ref{tab:galacticparams}, we calculated values of $U_\mathrm{c0}$ for NGC 4449 and DRC-8 using eqn. (\ref{eq:Uc0fromSFR}).
The mass densities listed in Table \ref{tab:galacticparams} have been calculated from the number densities or total masses and volumes quoted in the paper cited next to each mass density. The conversion we are utilizing for this is:
\begin{equation}
    \rho_\mathrm{ion} = 1.67\times10^{-24} n_\mathrm{ion} \hspace{1cm} \rho_\mathrm{gas} = 3.95\times10^{-24} n_\mathrm{gas}
\label{eq:massconvert}
\end{equation} 
assuming we have a He abundance of 10\% by number and all the hydrogen in molecular form and that most of the ions are protons (i.e. He is neutral). 

Further details about Table \ref{tab:galacticparams} are given in Appendix \ref{sec:galacticparam}. We note here, however, that the parameter
$\rho_0\Phi(R)/P_{c0}$ is very large in all cases, suggesting that all these galaxies are sub-Eddington unless
cosmic rays are transported by streaming and/or calorimetric with a star formation rate that declines relatively slowly with gas density, in which case more careful modeling is required.

\begin{table*}[t!]
	\centering
	\begin{tabular}{|c||c|c|c|c|c|c|c|c|c|c|}
		\hline
		Galaxy & $SFR^\mathrm{adv}$ & $U_\mathrm{c0}^\mathrm{adv}$ & $SFR^\mathrm{str}$ & $U_\mathrm{c0}^\mathrm{str}$ & $SFR^\mathrm{2Cal}$ & $U_\mathrm{c0}^\mathrm{2Cal}$ & $SFR^\mathrm{1.4Cal}$ & $U_\mathrm{c0}^\mathrm{1.4Cal}$ &  $SFR^\mathrm{obs}$ & $U_\mathrm{c0}^\mathrm{obs}$ \\
		\hline
		MW & $4.98\times10^{4}$ & $2.13\times 10^{6}$ & $76.0$ & $3251.2$ & $1.03\times10^{4}$ & $4.44\times10^{5}$ & $0.38$ & $16.1$ & $0.01$ & $10$ \\
		\hline
		M82 & $3.82\times 10^{4}$ & $1.64\times10^{6}$ & $63.6$ & $2733.0$ & $8.07\times10^{3}$ & $3.47\times10^{5}$ & $0.34$ & $14.5$ & $10$ & $525$ \\
		\hline
		LMC & $1049.8$ & $4737.8$ & $12.4$ & $56.1$ & $317.7$ & $1.43\times10^{3}$ & $0.31$ & $1.41$ & $0.4$ & $0.58$ \\
		\hline
		NGC 4449 & $783.7$ & $3.09\times10^{3}$ & $10.73$ & $42.26$ & $245$ & $965$ & $0.3$ & $1.19$ & $0.97$ & $3.82$ \\
		\hline
		DRC-8 & $2.54\times10^{9}$ & $2.17\times10^{7}$ & $1.78\times10^{6}$ & $1.53\times10^{4}$ & $4.74\times10^{8}$ & $4.05\times10^{6}$ & $4765$ & $40.8$ & $394$ & $3.18$\\
		\hline
	\end{tabular}
	\caption{First note the units for SFRs here are $(\rm M_{\odot} \unit yr^{-1})$ and that the units for $U_\mathrm{c0}$ are $(\rm eV \unit cm^{-3})$. The first and second columns are the calculated star formation rates and cosmic ray energy densities required to reach the Eddington limit for our different galaxy models for advection, while the third and fourth columns are the same for cosmic ray streaming. The fifth and sixth column, denoted as '2Cal', are the Eddington SFR and $U_\mathrm{c0}$ for the calorimetric limit where $\alpha = 2$, while the seventh and eighth column, denoted as '1.4Cal', are the values for the calorimetric limit where $\alpha = 1.4$. Lastly, we have pulled the observed star formation rates from Table \ref{tab:galacticparams} along with their calculated $U_\mathrm{c0}$'s from eqn. (\ref{eq:Uc0fromSFR}) and placed them in last two columns for easy comparison.}
	\label{tab:analytical_results}
\end{table*}

\begin{table}[]
	\centering
	
	\begin{tabular}{|c||c|c|c|}
		\hline
		Galaxy & $\rho_\mathrm{Edd}^\mathrm{adv}$ & $\rho_\mathrm{Edd}^\mathrm{str}$ & $\rho_\mathrm{0}^\mathrm{obs}$ \\
		\hline
		MW & $5.91\times 10^{-27}$ & $2.16\times10^{-25}$ & $1.27\times10^{-21}$ \\
		\hline
		M82 & $5.56\times10^{-25}$ & $1.47 \times 10^{-22}$ & $1.75\times10^{-21}$ \\
		\hline
		LMC & $1.34\times10^{-27}$ & $1.18\times 10^{-26}$ & $1.12\times10^{-23}$ \\
		\hline
		NGC 4449 & $1,0\times10^{-26}$ & $2.26\times10^{-25}$ & $8.3\times10^{-24}$ \\
		\hline
		DRC-8 & $1.23\times10^{-28}$ & $1.16 \times 10^{-27}$ & $4.0 \times 10^{-21}$ \\
		\hline
	\end{tabular}
	\caption{The Eddington gas densities in units of $(\unit g \unit cm^{-3})$ for each galaxy in the pure advection and streaming models compared to their calculated values from observations in the last column. Note that our method for calculating mass densities from observations is noted at the end of \S\ref{subsec:parameters}.}
	\label{tab:ana_densities}
\end{table}
\section{Results}
\label{sec:Results}

\subsection{Analytical Results}
\label{subsec:anaResults}

We can substitute the galactic parameters from \S\ref{subsec:parameters} into each system from \S\ref{subsec:transportModels} and determine if it can reach its Eddington limit. The Eddington star formation rates for advection and cosmic ray streaming are shown in Table \ref{tab:analytical_results}, along with the observed star formation rates for each galaxy that were quoted in Table \ref{tab:galacticparams}. The Eddington star formation rates were derived using eqn. (\ref{eq:PcEdd}) for advection, streaming, and the calorimetric model with $\alpha = 1.4$, whereas for $\alpha =2$ we used eqn. (\ref{eq:PEDDcal2}). The resulting derived $P_\mathrm{Edd}$ were then plugged into eqn. (\ref{eq:Uc0fromSFR}) ($U_\mathrm{c} = 3P_\mathrm{c}$) to solve for the star formation rate.

We can see that for all of the advection models, the star formation rate required to reach the cosmic ray Eddington limit is much larger than the actual star formation rate of all five galaxies. When we implement cosmic ray streaming, the Eddington value becomes closer to the observed value but is still much larger for all galaxies. This is qualitatively consistent with the discussion at the end of \S\ref{sec:analytical}.

The galaxy that gets the closest to its observed star formation rate is M82. We find that M82's Eddington star formation rate is about five times larger than its actual star formation rate, while the next closest, NGC 4449, would have to have a star formation rate about eleven times its observed value to be super-Eddington. Therefore, initially with just the transport models, while streaming allows the system to get closer to the Eddington limit, it looks extremely unlikely that the star formation rates of any of these galaxies are capped by their cosmic ray Eddington limit.

\begin{figure}[t!]
    \centering
    \includegraphics[width=0.5\textwidth]{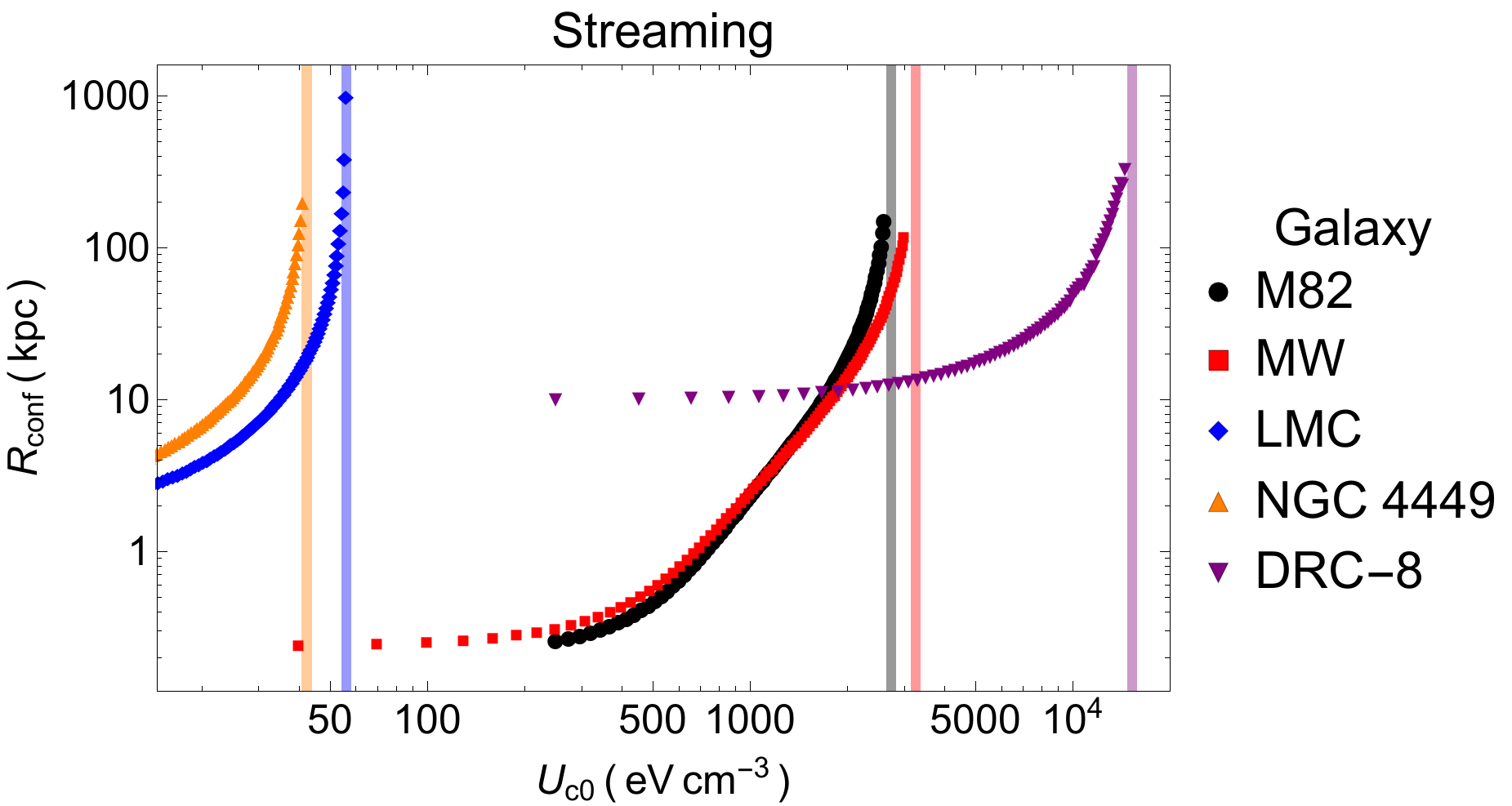}
    \includegraphics[width=0.5\textwidth]{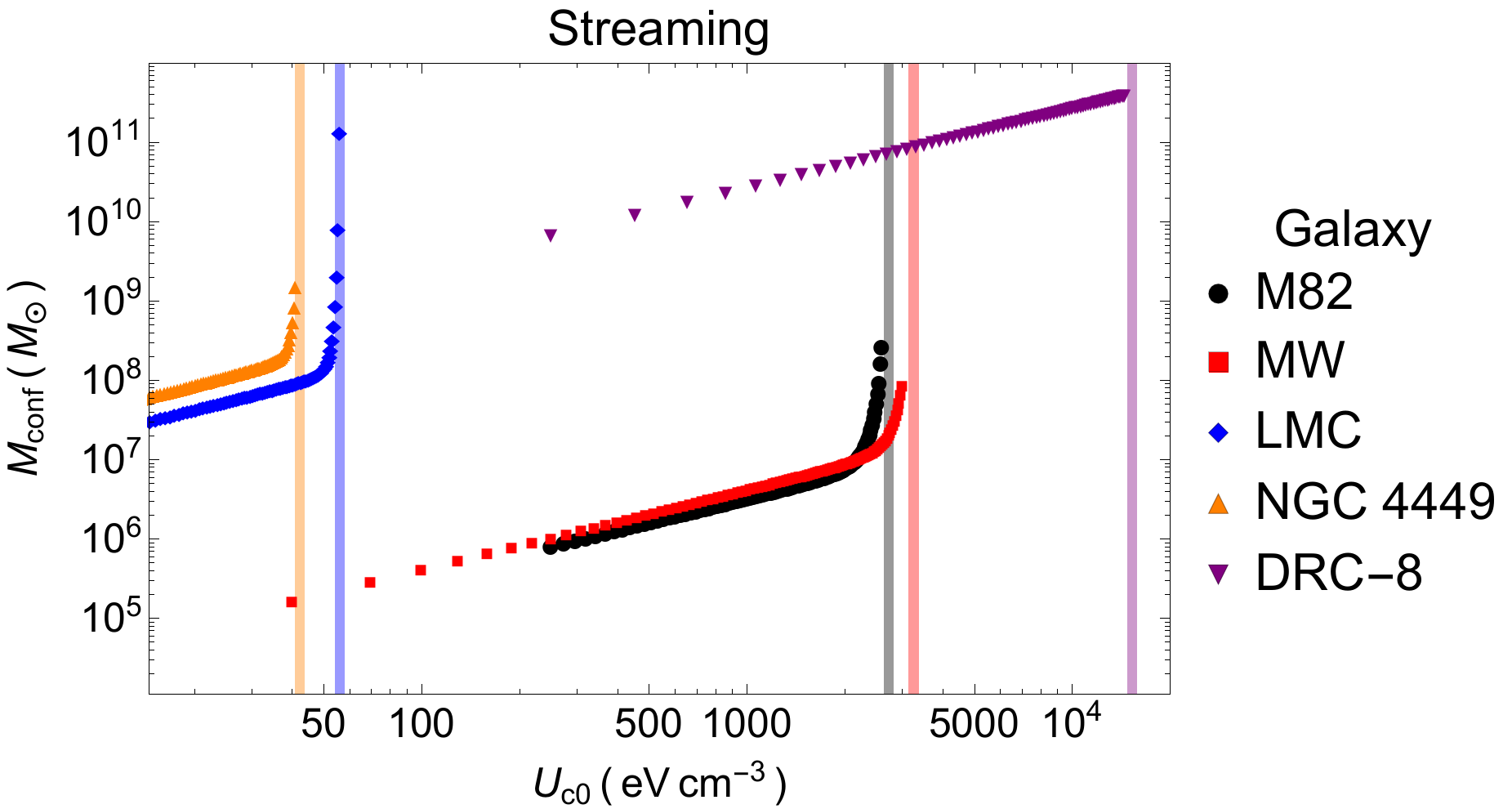}
    \includegraphics[width=0.5\textwidth]{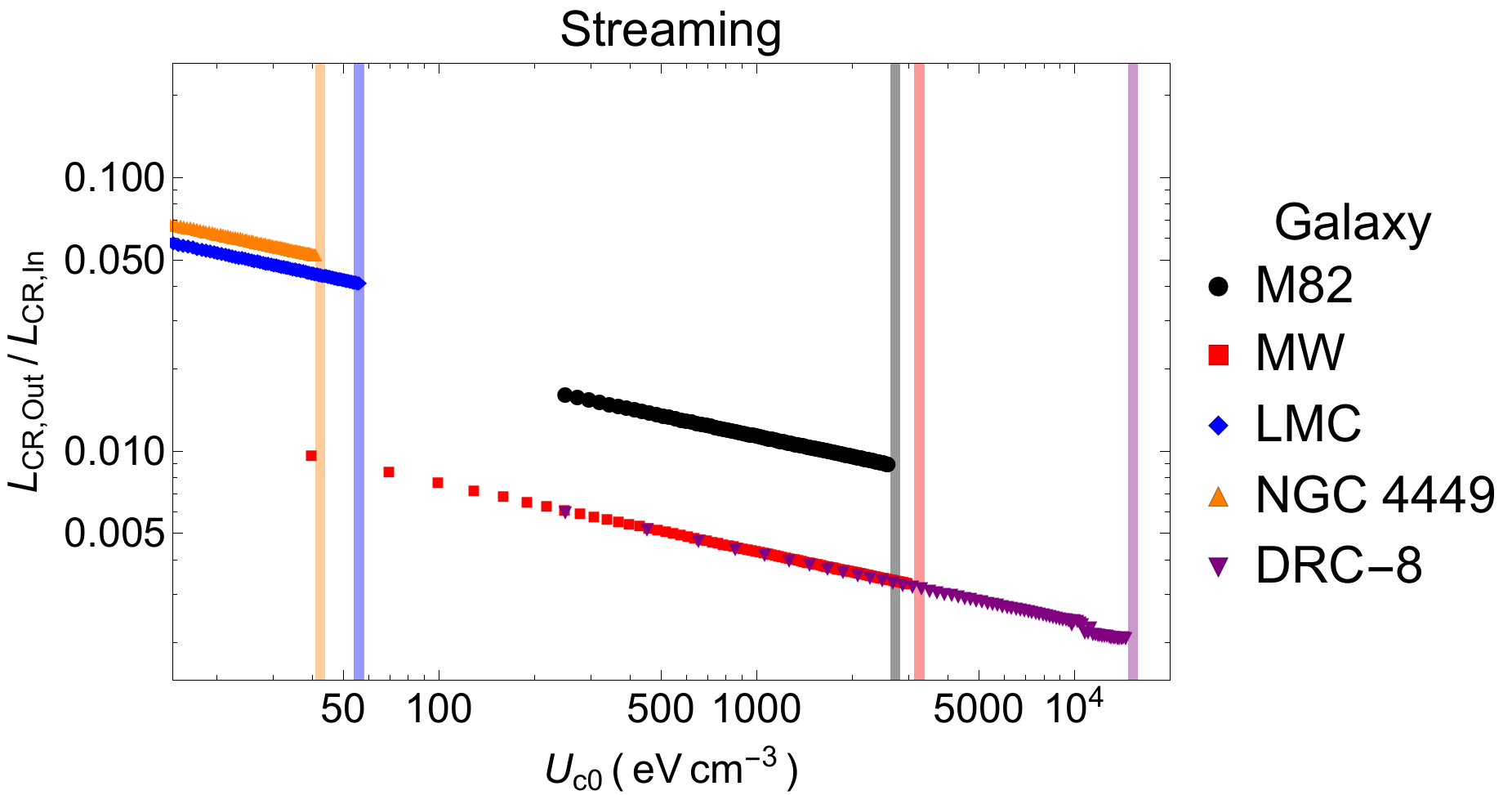}
    \caption{The radius of confinement, mass confined, and ratio of cosmic ray enthalpy luminosities in and out of our systems as functions of $U_\mathrm{c0}$ for each galaxy in our sample, assuming pure streaming (transport with $Q$ set to zero). The Eddington cosmic ray energy density derived from the star formation rates in Table \ref{tab:analytical_results} is marked with a vertical line for each galaxy.
    }
    \label{fig:CombinedUc0VaryPlotsJS}
\end{figure}

In Figure \ref{fig:CombinedUc0VaryPlotsJS}, we have plotted how the radius of confinement $R_{conf}$ and mass confined $M_{conf}$ (see eqn. (\ref{eq:Mconf})) change with $U_{c0}$ for the pure streaming model for all five galaxies.

It is also of interest to compare the cosmic ray enthalpy flux at the inner boundary of each galaxy with the enthalpy flux at the radius of confinement (where $P_\mathrm{c} = 10^{-15} \unit dynes \unit cm^{-2}$. For cosmic ray streaming, the enthalpy will show us the energy lost due to the heating of the interstellar gas. The removal of this energy from the cosmic rays is modeled in eqn. (\ref{eq:fulltransport}) as the $-v_\mathrm{A} \del P_\mathrm{c}$ term. We define the cosmic ray enthalpy flux as: 
\begin{equation}
    F_\mathrm{c} = v_\mathrm{A} (U_\mathrm{c} + P_\mathrm{c}) = v_\mathrm{A,0}P_c\lr{\frac{R}{r}}^2\lr{\frac{\rho_\mathrm{0}}{\rho}}^{1/2} \lr{\frac{\gamma}{\gamma - 1}}
\label{eq:enthalpyflux_general}
\end{equation}
where we have used $U_c=P_c/(\gamma -1)$ and have taken $B \propto 1/r^2$ to ensure $\del \cdot \mathbf{B} = 0$. 

\begin{figure}[t!]
    \centering
    \includegraphics[width=0.5\textwidth]{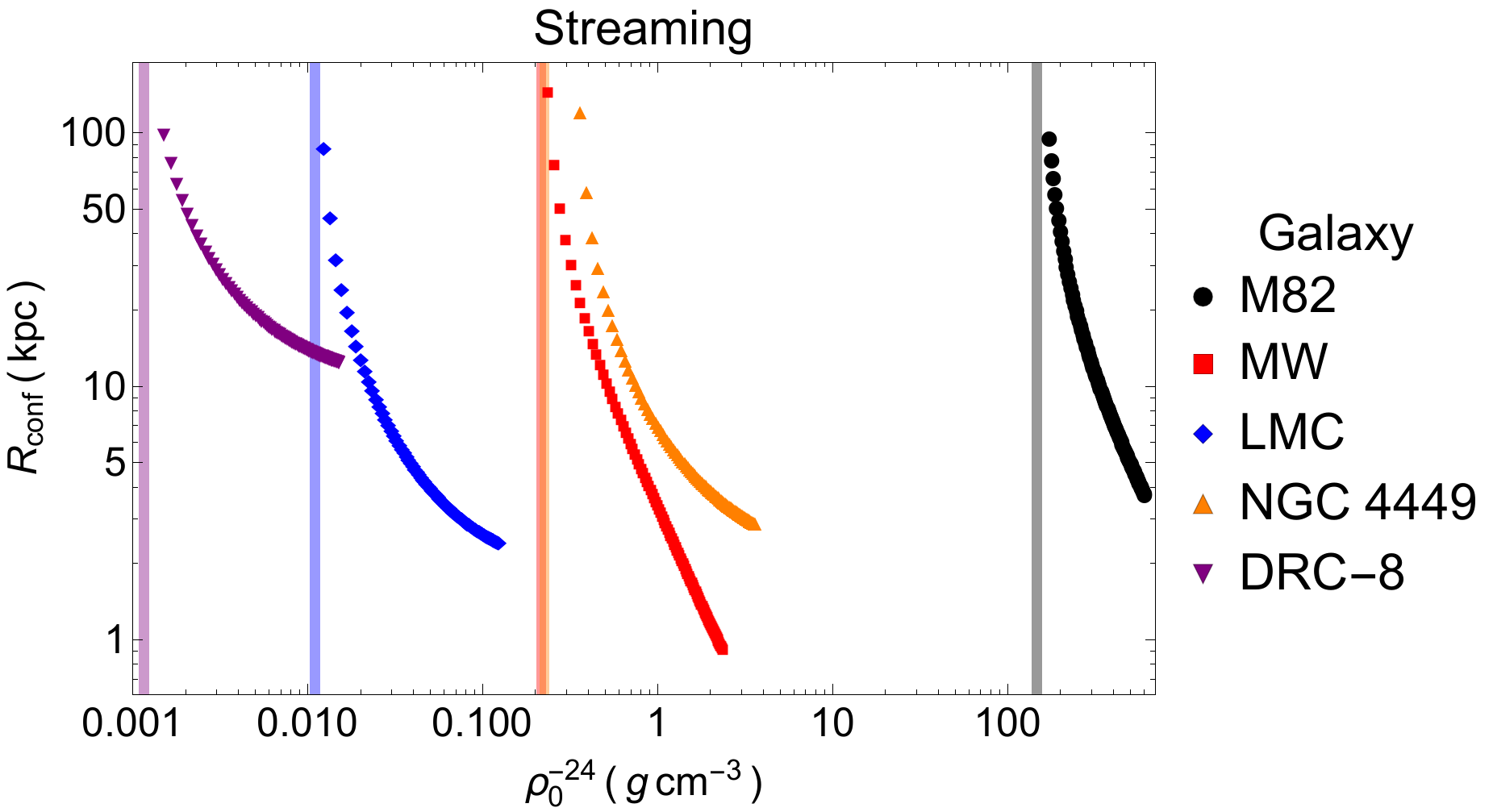}
    \includegraphics[width=0.5\textwidth]{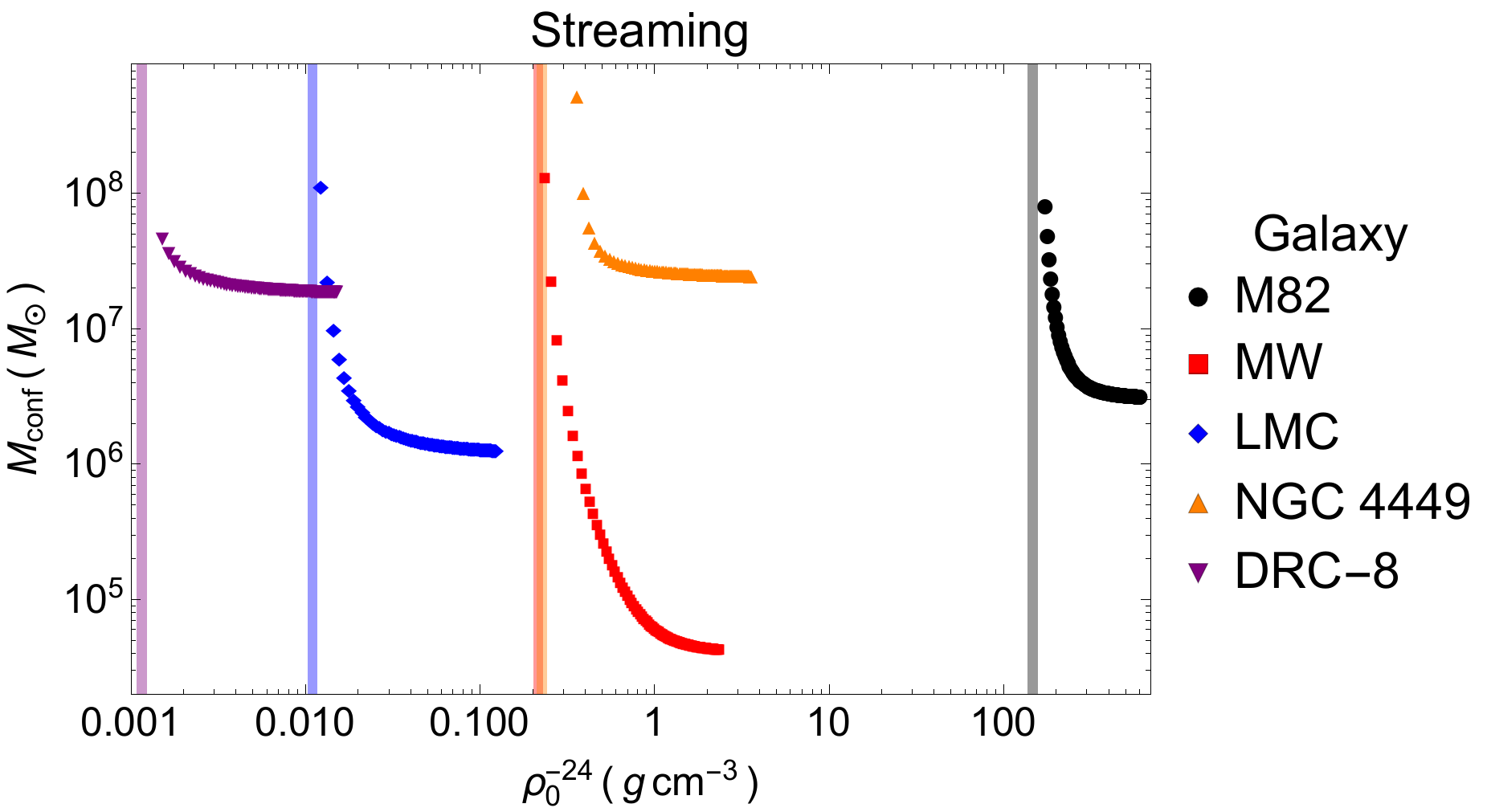}
    \includegraphics[width=0.5\textwidth]{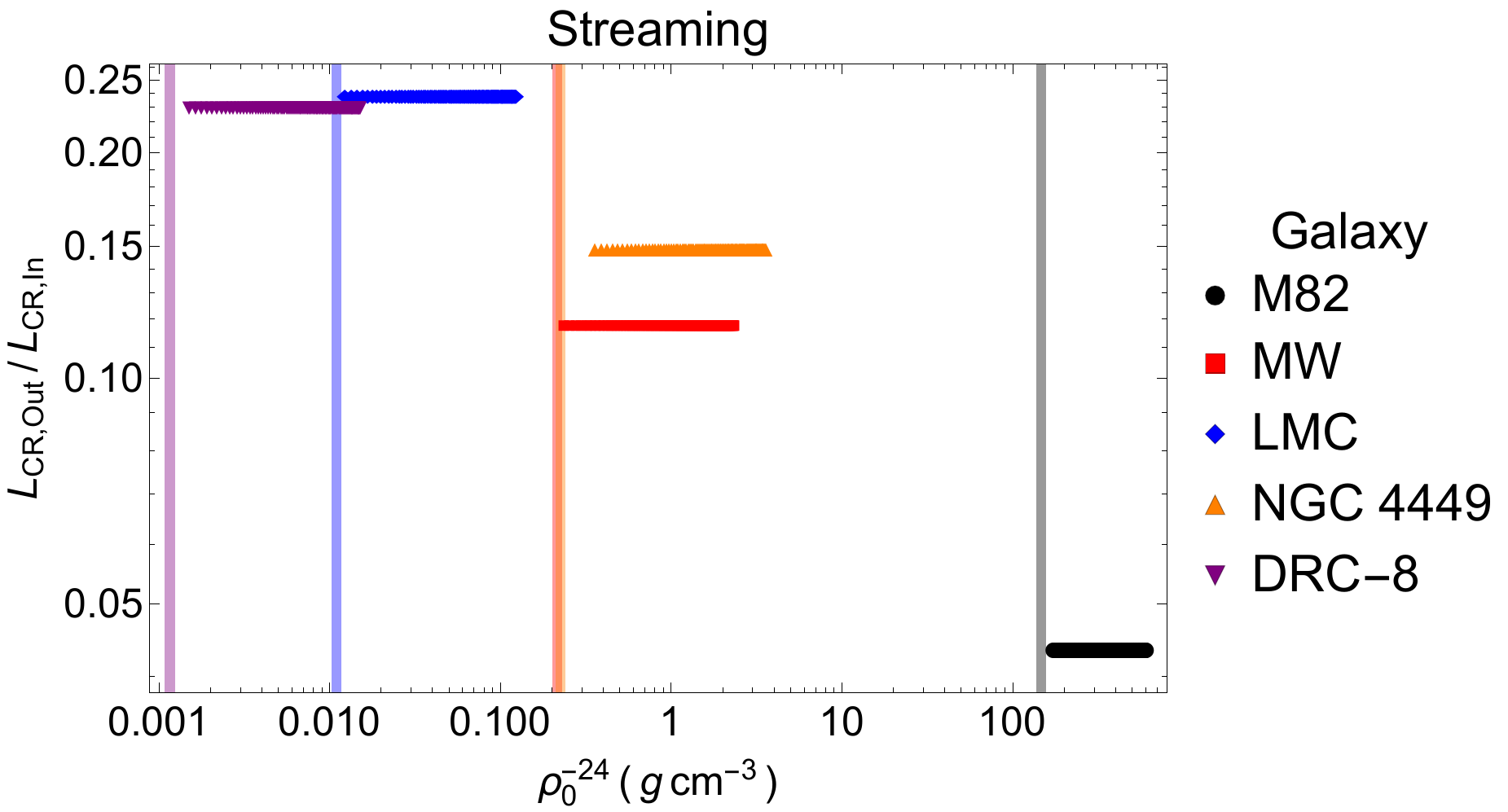}
    \caption{The radius of confinement, mass confined, and ratio of cosmic ray enthalpy luminosities as functions of $\rho_\mathrm{0}$ with the galaxy models assuming pure streaming transport with $Q$ set to zero and $U_\mathrm{c0}$ set to the value for each galaxy found in Table \ref{tab:galacticparams}. The Eddington gas densities from Table \ref{tab:ana_densities} are represented by vertical lines on each plot.}
    \label{fig:Combinedrho0VaryPlotsJS}
\end{figure}

We can then find the difference in the enthalpy luminosities at the inner boundary and the radius of confinement for each system. This difference can be derived from eqn. (\ref{eq:fulltransport}) without sources, losses, or diffusion, which we first rewrite as:
\begin{equation}
    \del \cdot \vec{F_\mathrm{c}} = \vec{v_\mathrm{A}} \cdot \del P_\mathrm{c}
\end{equation}
where we have substituted in eqn. (\ref{eq:enthalpyflux_general}). If we integrate both sides over the shell comprising the cosmic ray supported envelope, we find: 
\begin{equation}\label{eq:L}
L_h(R_{conf}) - L_h(R)= 4 \pi \int_R^{R_{conf}}r^2v_\mathrm{A}\cdot\nabla P_\mathrm{c} dr
\end{equation}
where $L_h(r) = 4 \pi r^2 F_\mathrm{c}(r)$ is the enthalpy luminosity. Therefore, eqn. (\ref{eq:L}) shows that the difference in enthalpy luminosities ($L_h$ here) at the confinement radius and at the base of our mass distribution equals the heat deposited. We can see from Fig. (\ref{fig:CombinedUc0VaryPlotsJS}) that essentially all the cosmic ray energy is expended between $R$ and $R_{conf}$. Although a detailed study of thermal balance is beyond the scope of this work, we note that collisionless cosmic ray heating has been argued to be important for warm, ionized extraplanar gas in the Milky Way \citep{WienerWIM2013}. To roughly estimate  the potential importance of heating, we define a heating timescale $\tau_h$ as the ratio of the confined mass thermal energy content to the cosmic ray enthalpy flux integrated over the area of the base:
\begin{equation}\label{eq:tauh}
\tau_h\equiv\frac{3M_{conf}k_BT}{32\pi mR^2P_\mathrm{Edd}v_{A0}},
\end{equation}
where $T$ and $m$ are the mean temperature of the atmosphere and mean particle mass, respectively. Taking parameters from Tables 1 and 2 and assuming $T=10^4$K, $m=1.0\times 10^{-24}$ g, we find $\tau_h\sim 5.0\times 10^5$ yr for the Milky Way and $2.1\times 10^5$ yr for NGC 4449. These short timescales suggest that even though $P_\mathrm{Edd}$ itself is much larger than any known value, a more modest cosmic ray pressure could heat the confined gas to the point that it provides additional thermal pressure, possibly leading to loss of hydrostatic equilibrium.

We can see from Figure \ref{fig:CombinedUc0VaryPlotsJS} that around the Eddington cosmic ray energy density (the vertical line in each plot), the radius of confinement and especially the mass confined values sharply turn vertical. Therefore, a reasonably accurate measure of where the cosmic ray Eddington limit is for each galaxy is to find the $U_\mathrm{c0}$ around where this asymptotic behavior begins. However, we want to caution that while the mass confined will be well-defined for values of $U_\mathrm{c0}$ far from the Eddington limit, as it gets closer to $U_\mathrm{Edd}$, very small changes in $U_\mathrm{c0}$ make large changes in $R_{conf}$, and so $M_{conf}$ will not be as well-defined near that boundary. 

One is able to derive a similar relationship for $\rho_\mathrm{Edd}$ as we do for $P_\mathrm{Edd}$ in eqn. (\ref{eq:PcEdd}). In this case, the Eddington density describes the maximum density for each model where the fixed star formation rate can still reach the Eddington limit and launch a wind. The Eddington gas densities for advection and streaming are shown in Table \ref{tab:ana_densities}.

We have again plotted the radius of confinement and mass of confinement but for a varying $\rho_\mathrm{0}$ in Figure \ref{fig:Combinedrho0VaryPlotsJS}. The value of $\rho_\mathrm{Edd}$ for each galaxy is shown as a vertical line for each galaxy on Figure \ref{fig:Combinedrho0VaryPlotsJS} where we have fixed our value of $U_\mathrm{c0}$ and vary the value of $\rho_\mathrm{0}$. Note that for these plots, the values of $\rho_\mathrm{0}$ chosen do not line up with the values for the galaxy listed in Table \ref{tab:galacticparams}. We have done this to ensure that we can find the Eddington limit for these galaxies, even if it occurs at extremely small values that are far below the calculated values for each galaxy.

\begin{figure*}[t!]
    \centering
    \includegraphics[width=0.49\textwidth]{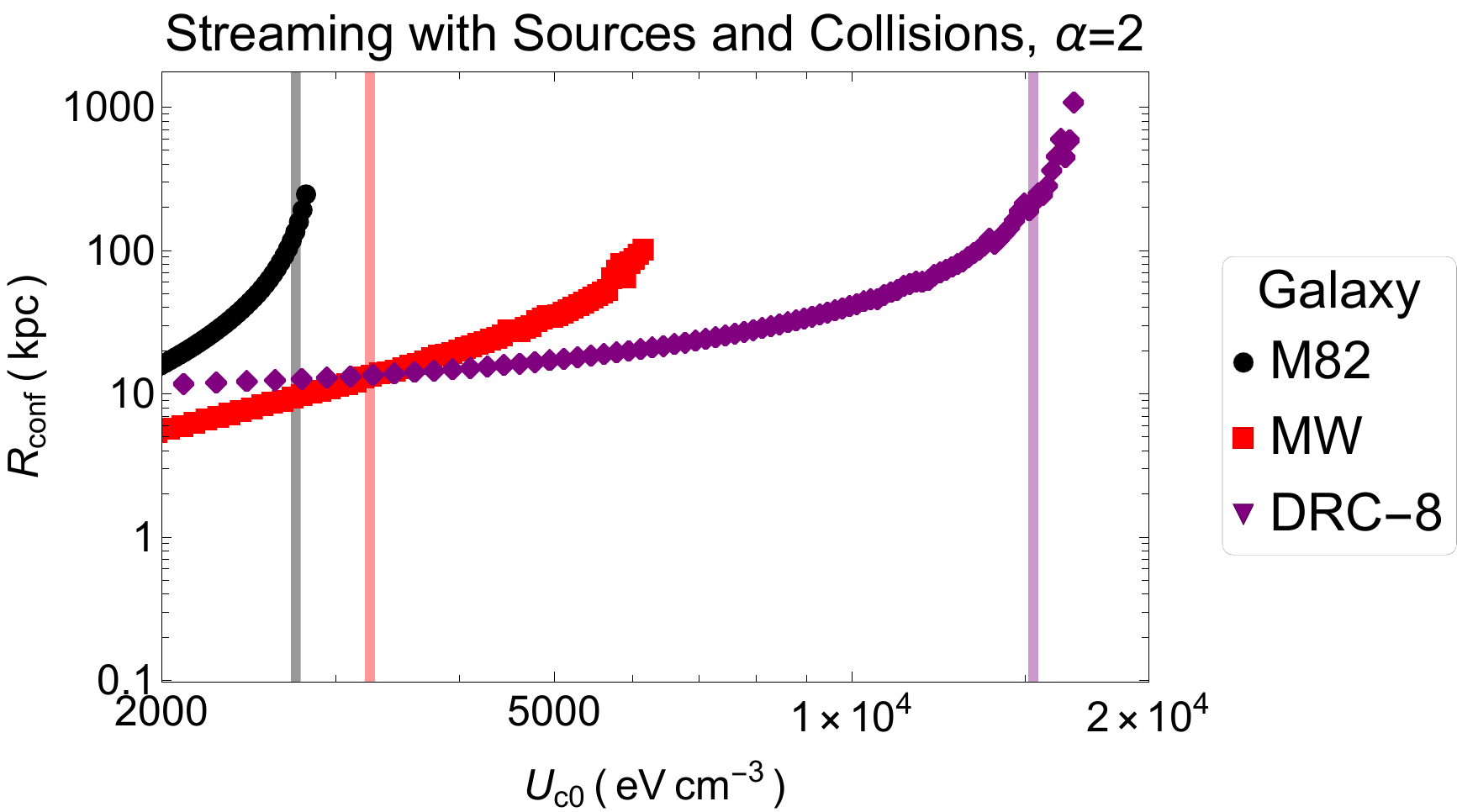}
    \includegraphics[width=0.49\textwidth]{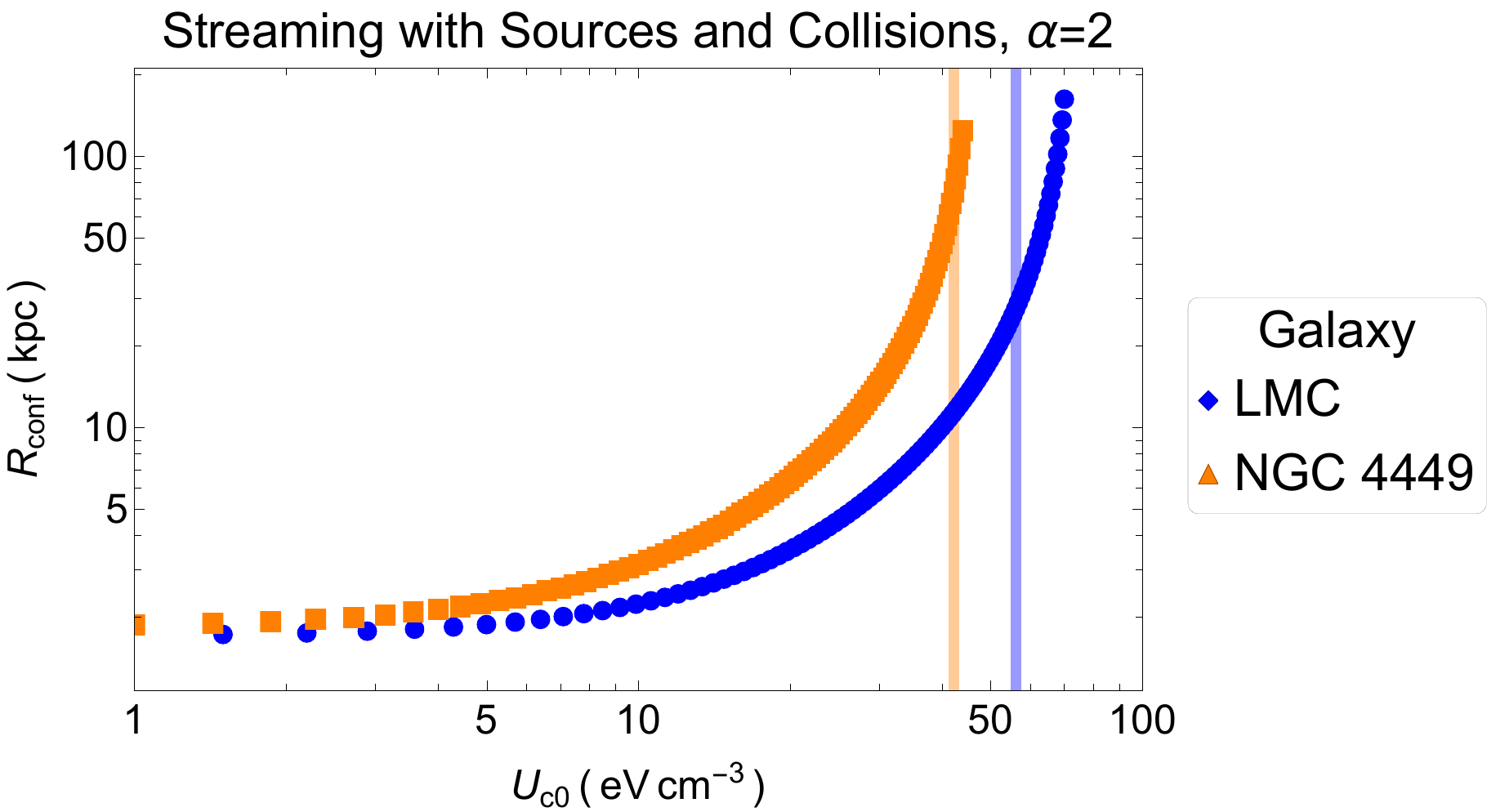}
    \includegraphics[width=0.49\textwidth]{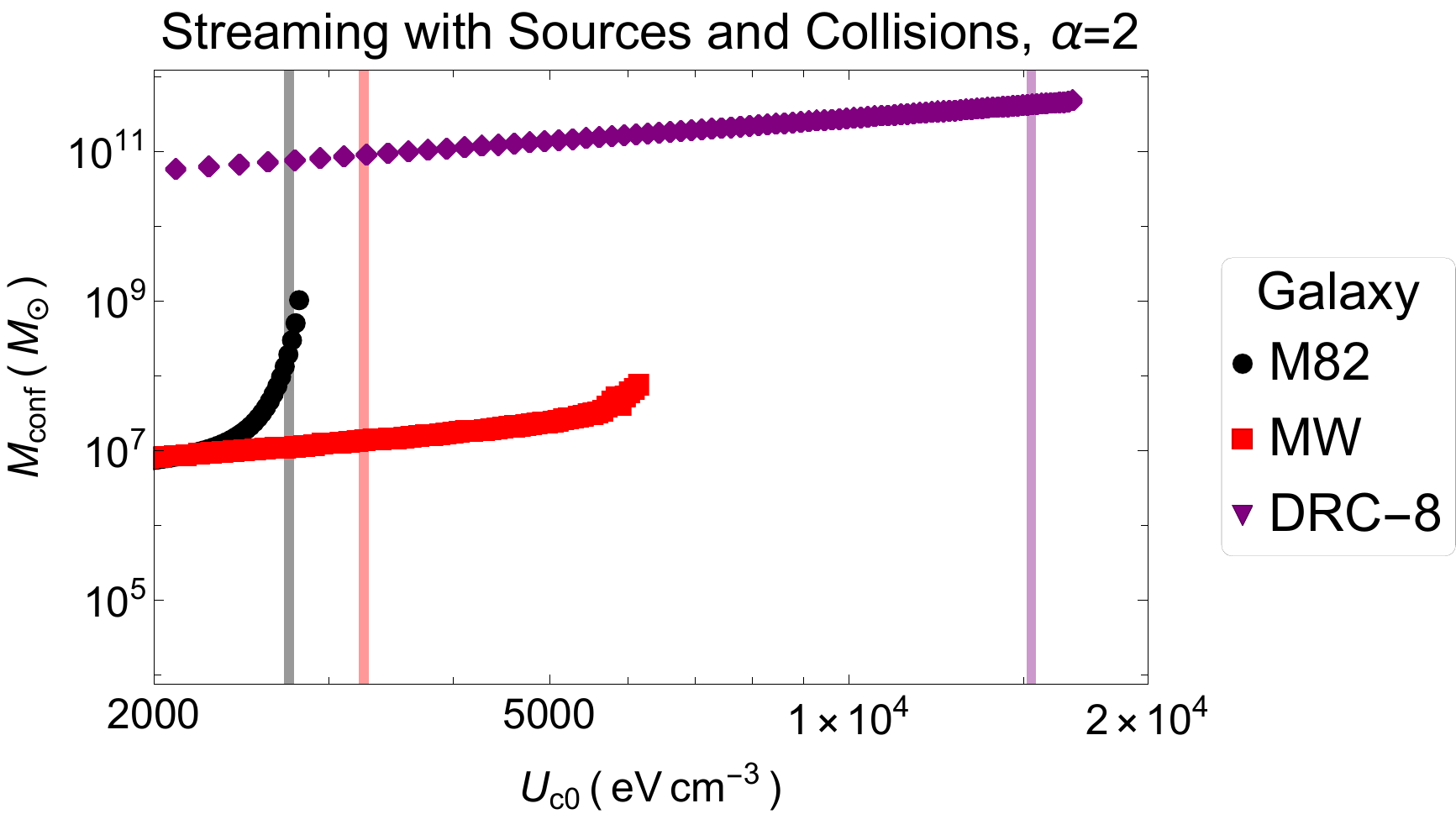}
    \includegraphics[width=0.49\textwidth]{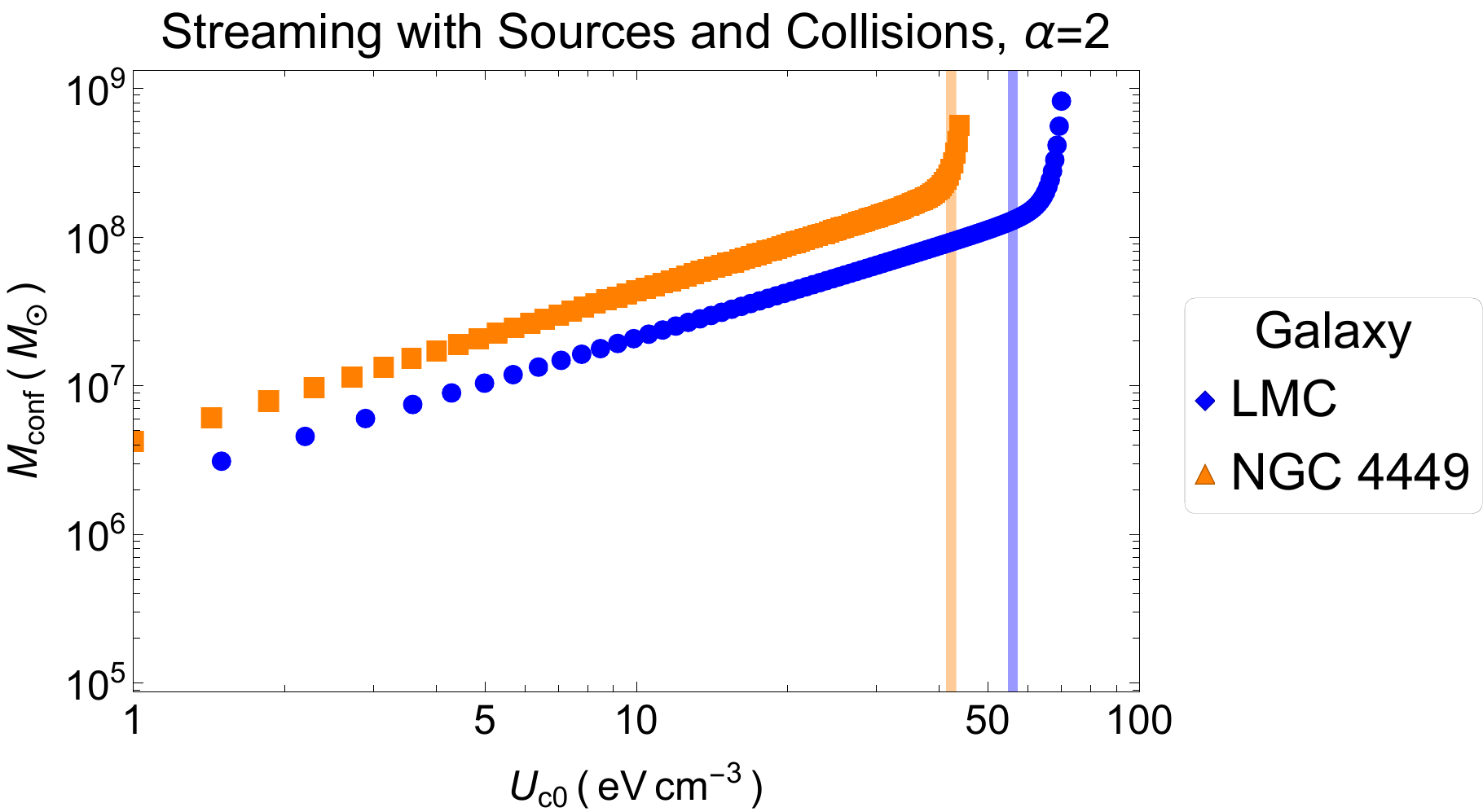}
    \includegraphics[width=0.49\textwidth]{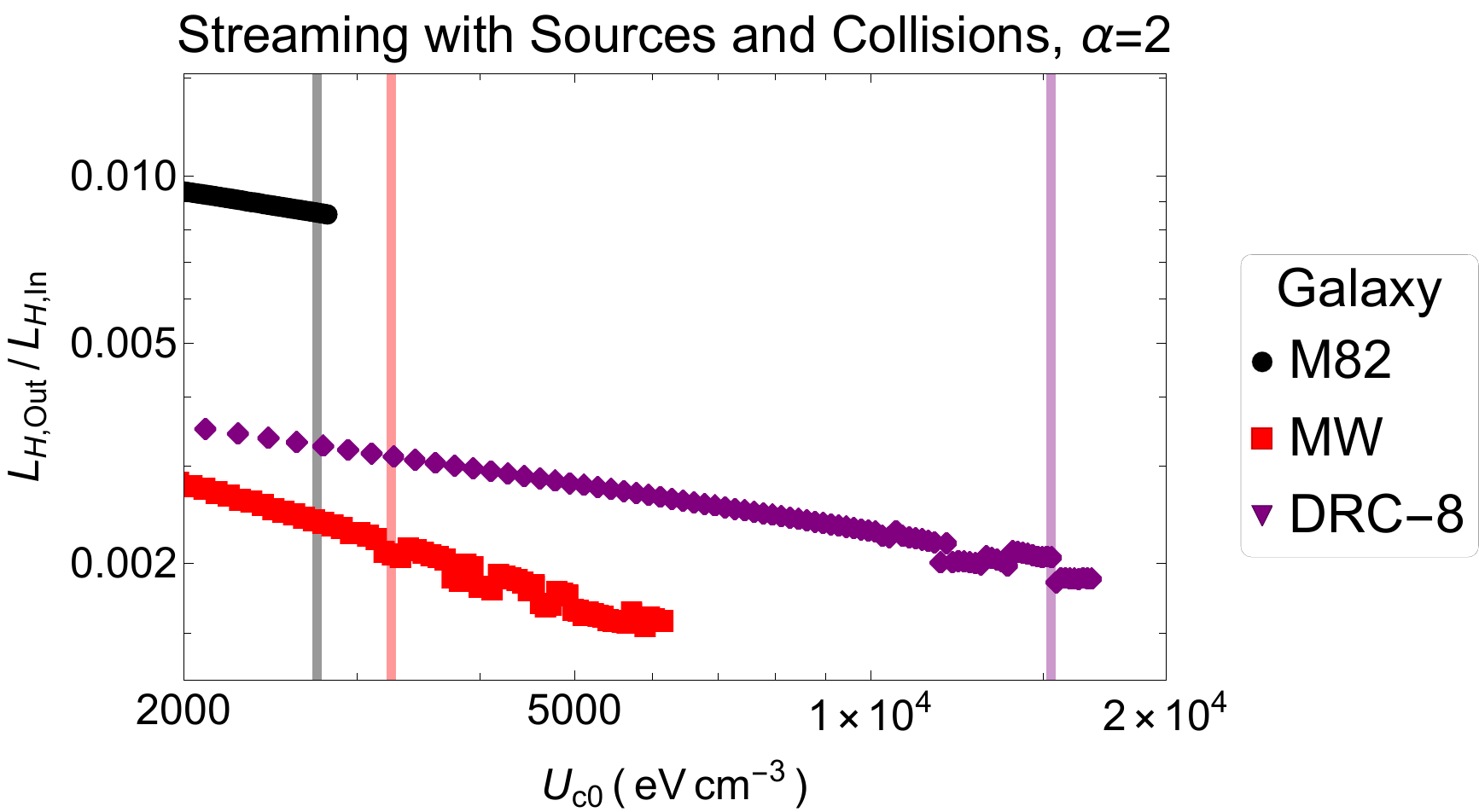}
    \includegraphics[width=0.49\textwidth]{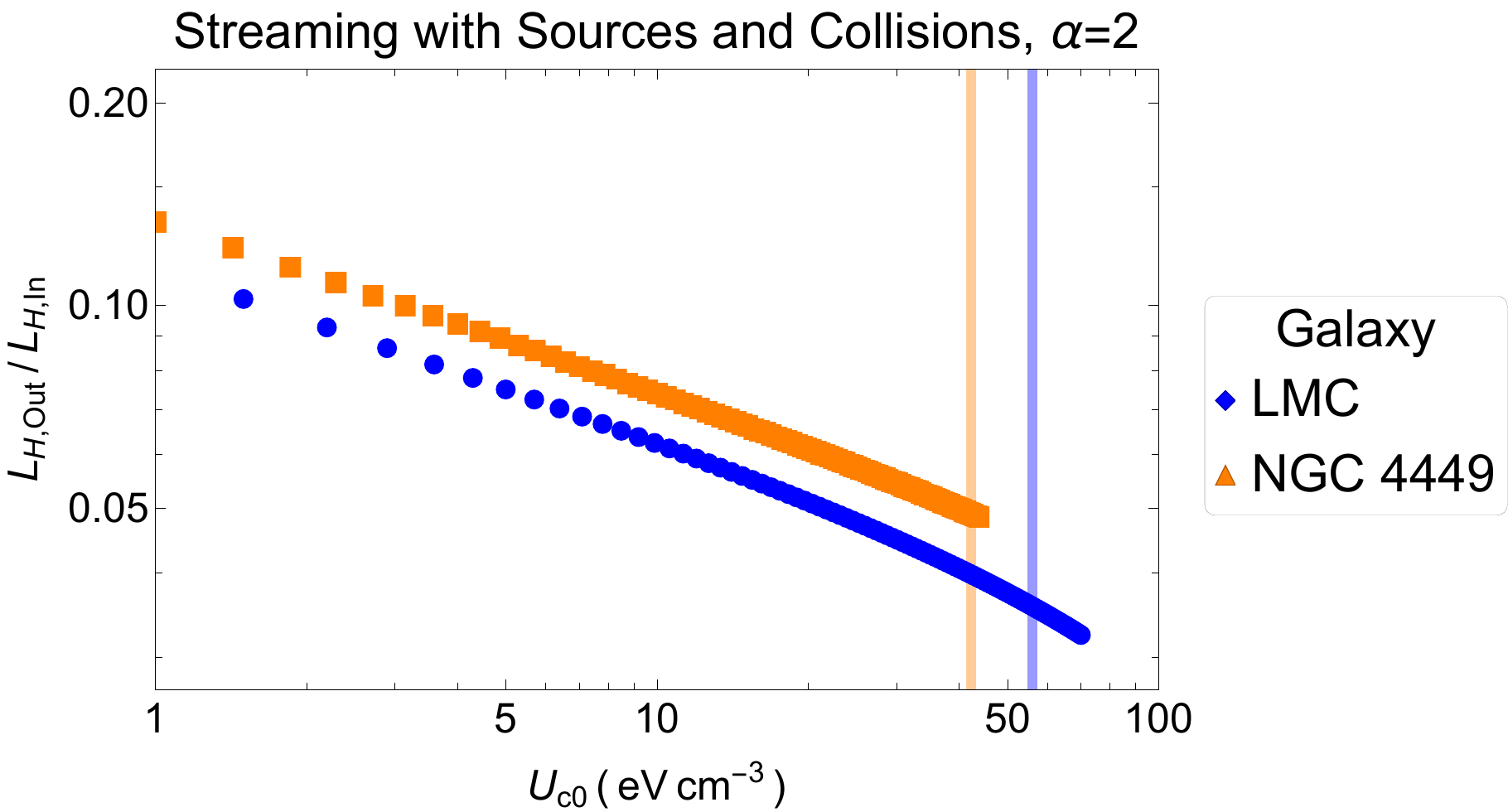}
    \caption{The radius of confinement, mass confined, and ratio of cosmic ray enthalpy luminosities into and out of our system as functions of $U_\mathrm{c0}$ for the galactic models with streaming, sources and collisions, assuming an $\alpha = 2$ K-S Law. The Eddington cosmic ray energy density for \emph{pure streaming} is marked with a vertical line for each galaxy in order to make easy comparisons.}
    \label{fig:CombinedUc0VaryPlotsSSL}
\end{figure*}

\begin{figure*}[t!]
    \centering
    \includegraphics[width=0.49\textwidth]{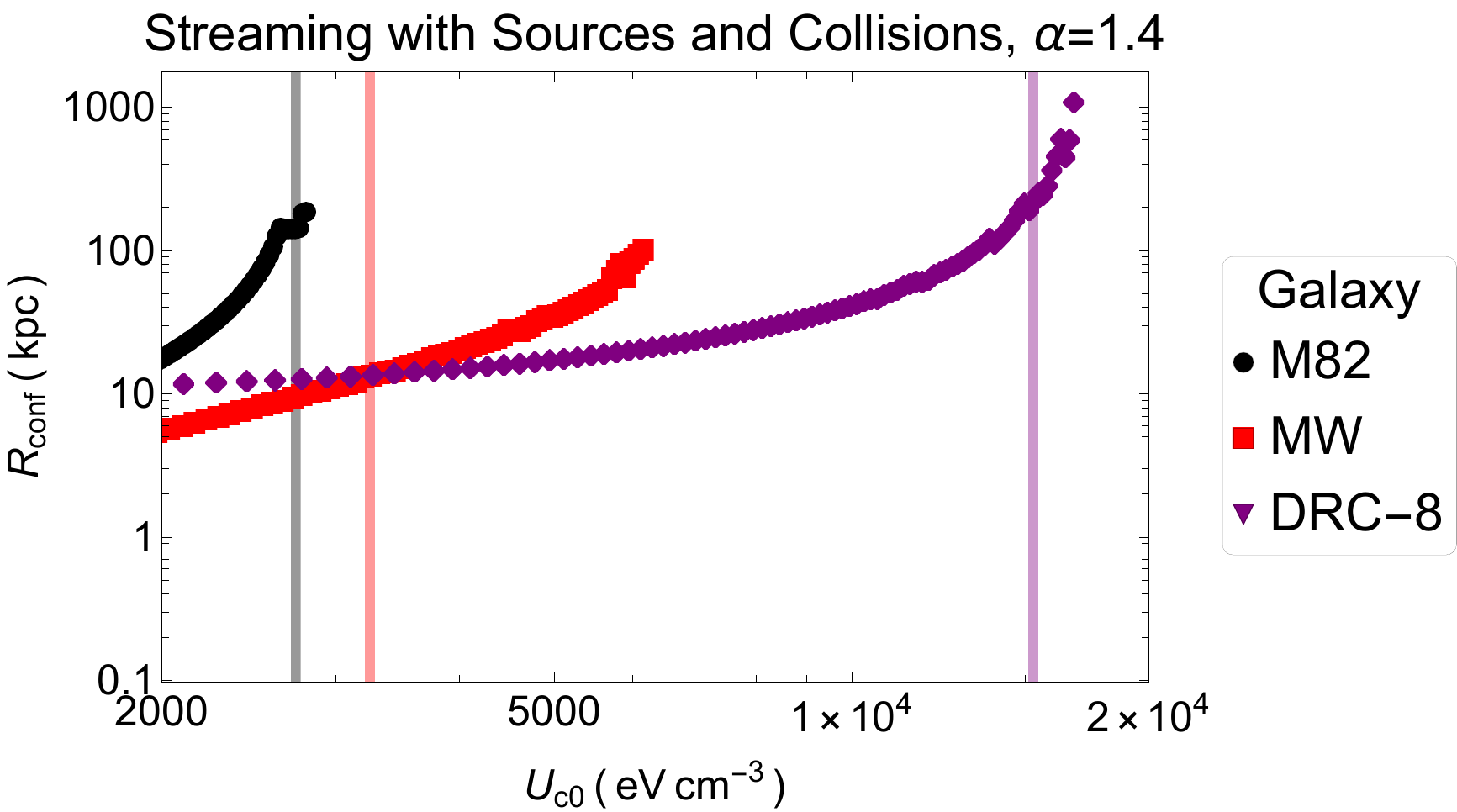}
    \includegraphics[width=0.49\textwidth]{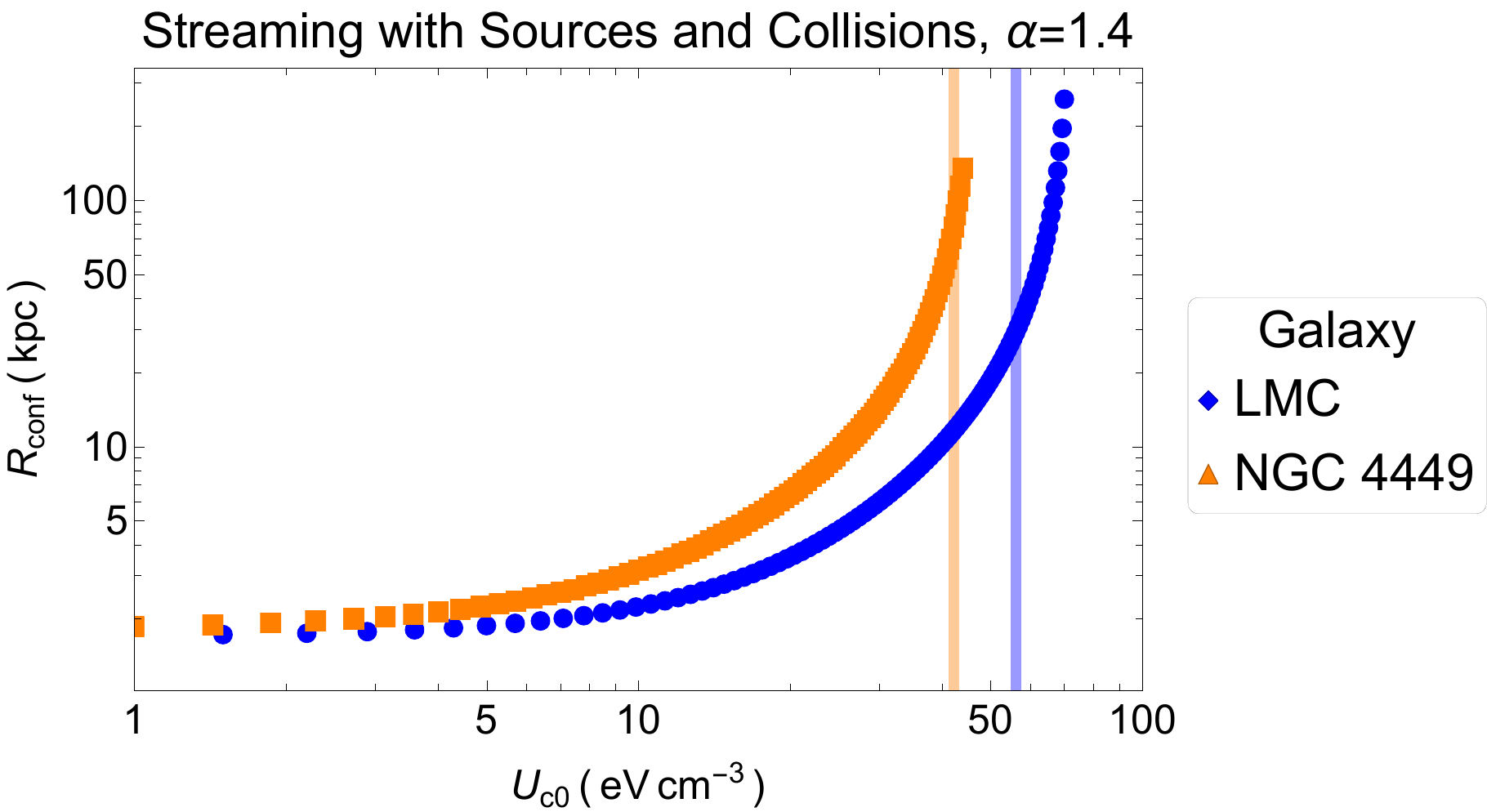}
    \includegraphics[width=0.49\textwidth]{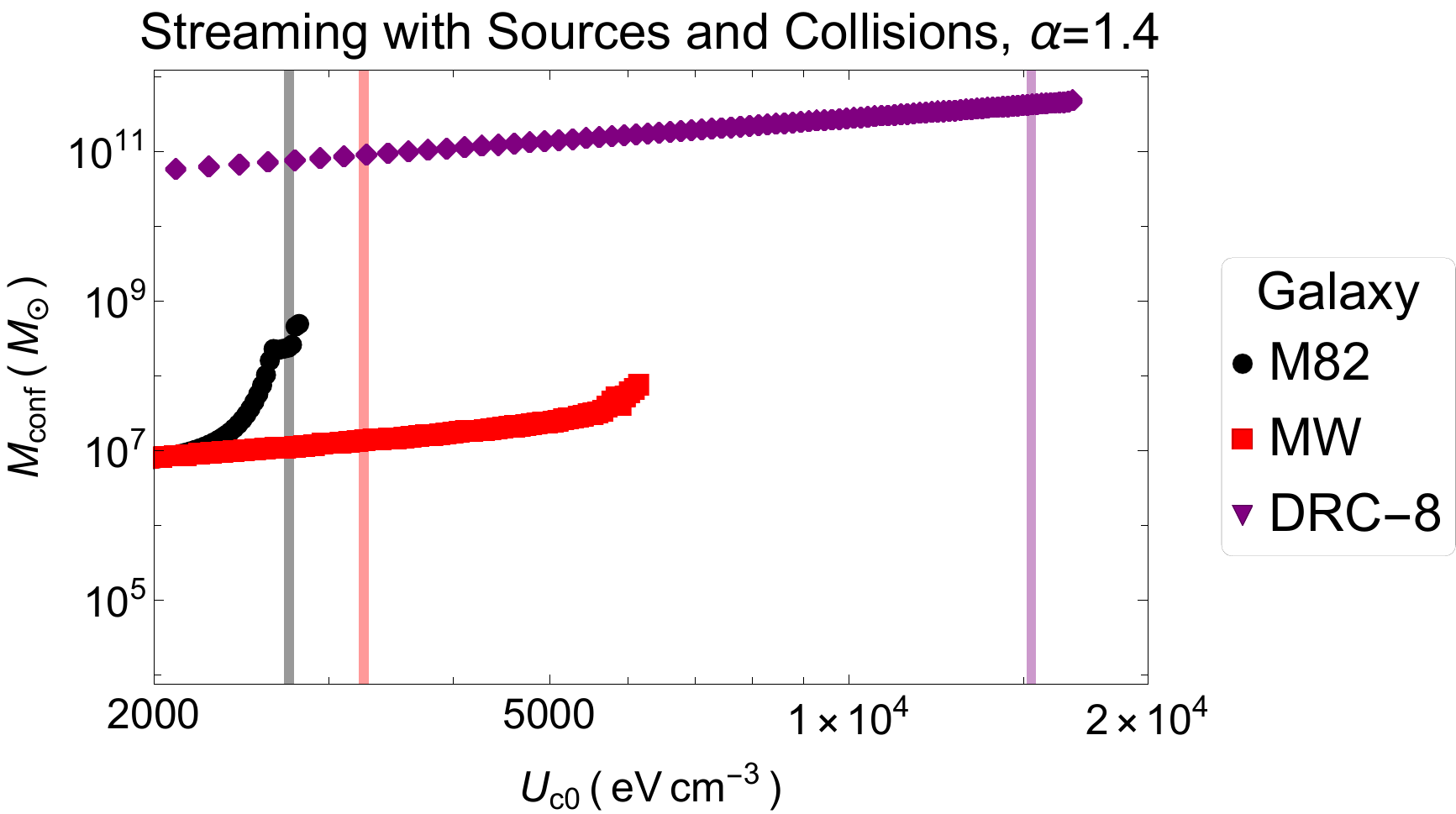}
    \includegraphics[width=0.49\textwidth]{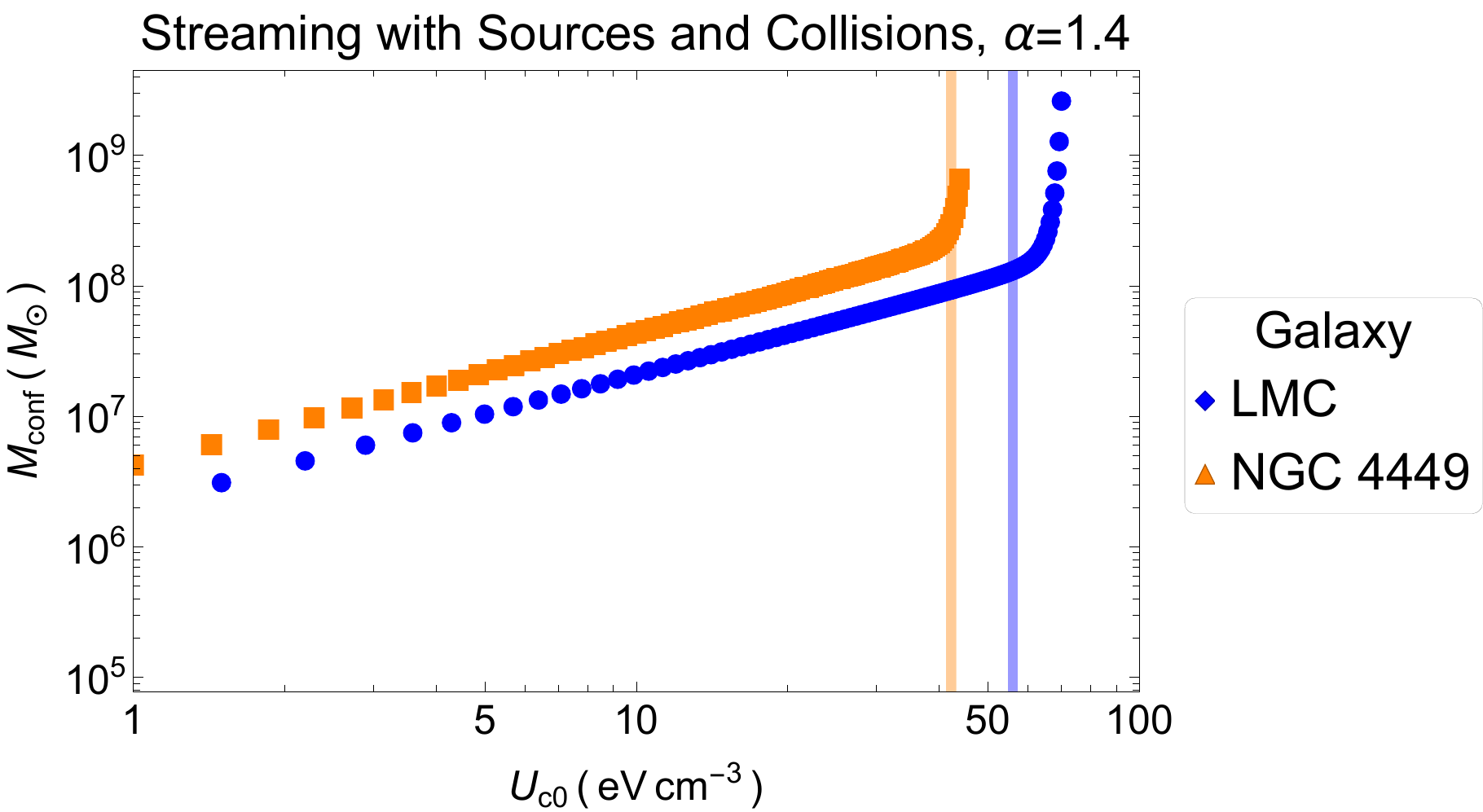}
    \includegraphics[width=0.49\textwidth]{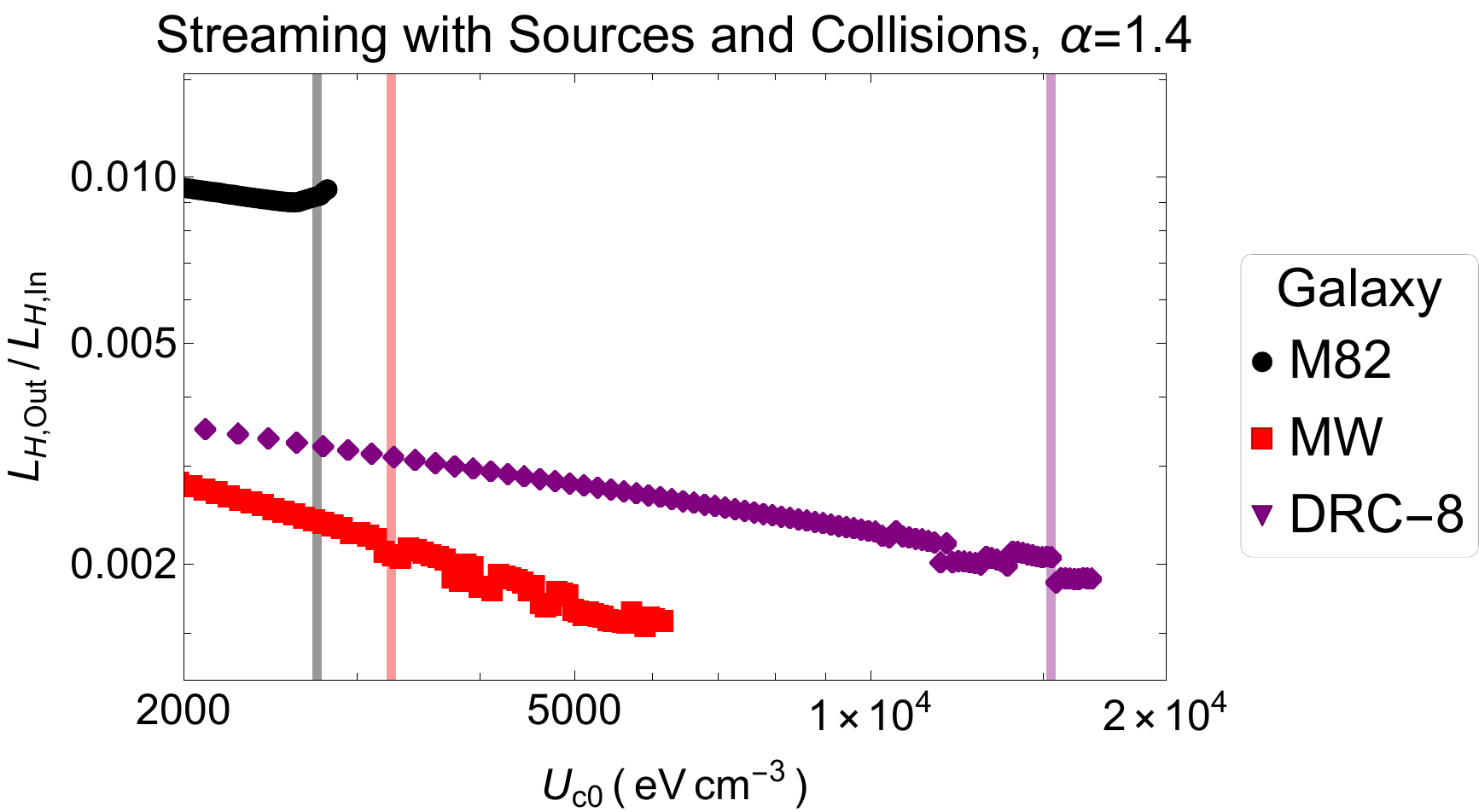}
    \includegraphics[width=0.49\textwidth]{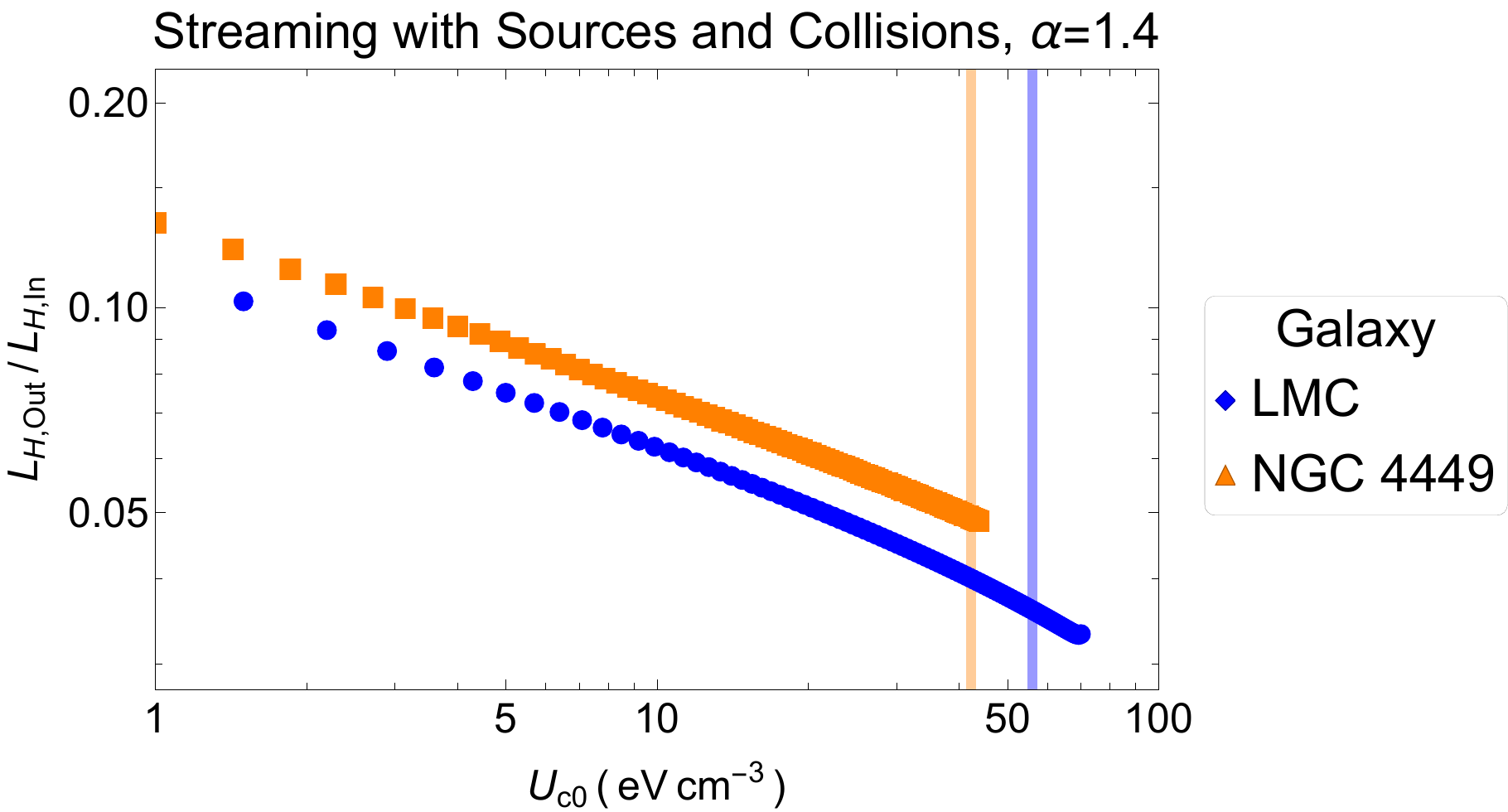}
    \caption{The radius of confinement, mass confined, and ratio of cosmic ray enthalpy luminosities into and out of our system as functions of $U_\mathrm{c0}$ for the galactic models with streaming, sources and collisions, assuming an $\alpha = 1.4$ K-S Law. The Eddington cosmic ray energy density for \emph{pure streaming} is marked with a vertical line for each galaxy in order to make easy comparisons.}
    \label{fig:CombinedUc0VaryPlotsSSL_LowerSFR}
\end{figure*}

We can see a similar behavior in these plots as with the varying $U_\mathrm{c0}$ plots where around the value of $\rho_\mathrm{Edd}$, the lines turn vertical and approach an asymptotic value.
Similarly to what we found when we varied $U_\mathrm{c0}$, none of the five galaxies have observed densities that fall below their Eddington density. The two galaxies that come the closest to reaching their Eddington gas density are NGC 4449 and M82 with both about or a little over an order of magnitude away from their Eddington values.

From this analysis, we can see then that there are two ways to reach the Eddington limit for cosmic rays. A galaxy either needs to have a large enough star formation rate that it can break hydrostatic equilibrium or have a gas density that is low enough that it requires very little energy injected into cosmic rays to launch a wind. However, since we are generally interested in the star formation rate and amount of cosmic ray injection needed to launch a wind, we will focus on just varying $U_\mathrm{c0}$ for the rest of this work.

Now that we have analyzed the pure transport model results, we will add in sources from star formation and losses from hadronic collisions to determine their effect on our system with cosmic ray streaming.

\subsection{Numerical Results}
\label{subsec:numResults}
Now that we have calculated the Eddington limit for simple systems, we can expand our analysis to include the sources and collisions and observe their effects on these galactic models. To make this easier for us when calculating, we non-dimensionalize our equations to the base values as outlined in Appendix \ref{sec:nondimens}. We then vary the value of $U_\mathrm{c0}$ to determine the point at which the system becomes super-Eddington. However, since these models have no analytical form, and based on the solutions do not seem to asymptote to a specific value, we need to use a more empirical method to determine if the Eddington limit is reached. As \cite{TumlinsonCGM2017} note, a specific point for the boundary between the CGM and the ISM is not well-defined but \cite{HopkinsCRGalaxyForm2020} explain in a footnote of their work that typically, the CGM is taken to begin somewhere between $10-30 \unit kpc$ and ends at its virial radius which can fall anywhere between $200 - 400 \unit kpc$. This is a bit of an arbitrary range and so to attempt to be as accurate as possible, we will instead find the value for $U_\mathrm{c0}$ at which the Eddington curve turns upward, as we saw in Figure \ref{fig:CombinedUc0VaryPlotsJS}. 

For our description of cosmic ray sources, we will assume two different K-S Laws, one for which $\alpha = 2$ in eqn. (\ref{eq:ksLaw}) and another for which $\alpha = 1.4$. The results for $\alpha = 2$ are shown in Figure \ref{fig:CombinedUc0VaryPlotsSSL}, while the $\alpha = 1.4$ results are shown in Figure \ref{fig:CombinedUc0VaryPlotsSSL_LowerSFR}. Note that the vertical lines in each plot show where $U_\mathrm{Edd}$ for the \emph{pure streaming} case occurred. 

We can see that in general the addition of sources and collisions has an effect on the cosmic ray Eddington limit of a galaxy. Sources tend to reduce the Eddington limit while losses increase it. By construction, sources dominate losses  at the inner boundary $r=R$, where $S=U_{c0}/\tau_L > U_{c0}/\tau_C$, but decline more steeply with radius due to their higher dependence on density $\rho^2$ or $\rho^{1.4}$ vs $\rho$, so losses dominate sources in the bulk of the domain. While the  streaming term also declines with $r$, it dominates both sources and losses, due to its $\rho^{-1/2}$ dependence and relatively slow geometrical decline of $B$ with radius. Due to collisions dominating near the inner boundary, we find, especially for non-starburst galaxies, that the Eddington limit is pushed to larger values of $U_{c0}$ to counteract collisions taking energy away from the cosmic ray population.

We also find that the choice of our exponent in the K-S Law seems to almost make no difference in the overall effect of sources and collisions. The only galaxies that exhibit only minor shifts based on the K-S Law are the LMC and M82. For the LMC, we can see when comparing Figure \ref{fig:CombinedUc0VaryPlotsSSL} to Figure \ref{fig:CombinedUc0VaryPlotsSSL_LowerSFR} that its confined radii and mass are slightly larger for the final value of $U_{c0}$ when $\alpha = 1.4$. M82's change is even more difficult to see but a close analysis of the shape of the curves in both plots show that their two shapes are slightly different.

\section{Summary and Conclusions}
\label{sec:discuss}

In this paper, we explored the feasibility of a cosmic ray Eddington limit based on the important idea first proposed by \cite{socrateseddington2008} but reformulated to reflect current understanding of the circumgalactic environment. Our framework was based on E.N. Parker's argument for the existence of the solar wind: that the solar corona cannot be in hydrostatic equilibrium because its pressure asymptotes to a value much larger than the interstellar pressure. We applied this argument by developing a family of models of the ISM in which thermal gas is supported solely by cosmic ray pressure $P_c$ and searching for conditions under which these models have an asymptotic pressure greater than the circumgalactic pressure $P_\mathrm{CGM}$.

\begin{table*}[t!]
    \centering
    \begin{tabular}{|c||c|c|c|c|c|c|}
    \hline
         Galaxy & $U_\mathrm{c0}^\mathrm{adv}$ & $U_\mathrm{c0}^\mathrm{str}$ & $U_\mathrm{c0}^\mathrm{2Cal}$ & $U_\mathrm{c0}^\mathrm{1.4Cal}$ & $U_\mathrm{c0}^\mathrm{SSC}$ & $U_\mathrm{c0}^\mathrm{obs}$ \\
         \hline
         MW & $2.13\times 10^{6}$ & $3251.2$ & $3.7\times 10^{4}$ & $16.1$ & $5700$ & $10$ \\
         \hline
         M82 & $1.64\times10^{6}$ & $2733.0$ & $4.1\times 10^{4}$ & $14.5$ & $2700$ & $525$ \\
         \hline
         LMC & $4737.8$ & $56.1$ & $1434$ & $1.41$ & $65$ & $0.58$ \\
         \hline
         NGC 4449 & $3.09\times10^{3}$ & $42.26$ & $9.1$ & $1.19$ & $41$ & $3.82$ \\
         \hline
         DRC-8 & $2.17\times10^{7}$ & $1.53\times10^{4}$ & $1.46\times 10^{5}$ & $40.8$ & $1.67\times 10^4$ & $3.18$\\
    \hline
    \end{tabular}
    \caption{The Eddington values for $U_\mathrm{c0}$ in units of $\rm eV \unit cm^{-3}$ for all transport models and galaxies in this paper. We have combined the two streaming with sources and collisions cases ($\alpha = 2$ and $\alpha = 1.4$) as it appears that the K-S Law's exponent has no discernible effect on the results of adding sources and collisions. Here, SSC just stands for streaming with sources and collisions.}
    \label{tab:comprehensive_results}
\end{table*}

We have provided a comprehensive summary of all the Eddington $U_\mathrm{c0}$'s for each galaxy in each different transport model in Table \ref{tab:comprehensive_results}. For the cases of streaming with sources and collisions, we have estimated the Eddington $U_\mathrm{c0}$ by choosing the point at which specifically the mass confined begins to asymptote vertically. We showed in Figure \ref{fig:CombinedUc0VaryPlotsJS} that this asymptotic behavior is a good indicator of having reached the Eddington limit.

Details of the models - geometry, galactic gravitational potential, modes of cosmic ray transport, sources due to star formation, and losses due to hadronic collisions - are described in \S\ref{sec:analytical}.  We considered three transport mechanisms:  self confinement due to Alfvenic streaming, diffusion with constant diffusivity $\kappa$, and the limit $\kappa\rightarrow 0$, which we termed the advection model, because the cosmic rays are essentially frozen to the gas and would be advected by any gas flow that was present.  For sources, we adopted a Kennicutt-Schmidt law, leading to a cosmic ray injection rate proportional to a power of the thermal gas density $\rho$.

We found several limiting cases in which $P_c$ is related to $\rho$ by a polytropic equation of state, for which hydrostatic equilibrium models are easily constructed, and which share a universal figure of merit: $\rho_0\Phi(R)/P_{co}$, the depth of the gravitational well confining the system in units of pressure. These models  give rise to an Eddington limit in the sense that for given values of $P_c$ and $\rho$ at a fiducial base radius $R$, the pressure either drops to $P_\mathrm{CGM}$ at some confinement radius $R_{conf}$, in which case the model is sub-Eddington, or never falls to $P_\mathrm{CGM}$ - super-Eddington. Thus, at least for the simplifying assumptions satisfied by our models, the concept of an Eddington limit for cosmic rays is well founded. From the polytropic solutions, models with $n > 0$ ($a > 1$) have $P_\mathrm{Edd}\sim \rho_0\vert\Phi(R)\vert$, while models with $n < 0$ ($a < 1$)have $P_\mathrm{Edd}\sim P_\mathrm{CGM}{-1/n}(\rho_0\vert\Phi(R)\vert)^{1+1/n}$. Based on this finding we conclude that  Alfvenic streaming, the only transport model which leads to a negative polytropic index, is the most likely candidate for reaching the Eddington limit. The calorimetric limit with star formation rate SFR$\propto\rho^{1.4}$ also leads to a model with $n < 0$, but this limit is only reached for extremely large gas densities and star formation rates.

Our models are very general, and so for concreteness, we picked five different galaxies that we believe are representative of the many types of star forming galaxies in our universe: a large spiral (the Milky Way), a gas rich dwarf (the LMC), a large starburst (M82), a dwarf starburst (NGC 4449) and a large, gas rich, dusty galaxy viewed at $z\approx 4$ (DRC-8). Their parameters are given in Table 1.

We found that advection as a method of cosmic ray transport is not capable of reaching a cosmic ray Eddington limit, almost always requiring a star formation rate 3 or more orders of magnitude larger than any actual observed SFR. Diffusion at constant diffusivity turns out to be a poor model for reaching the cosmic ray Eddington limit. In order for a model with cosmic ray diffusion as the primary mode of transport, one of two things must occur. One way is that the density of the galaxy must be extremely low ($\approx 10^{-27} \unit g \unit cm^{-3}$), down to a level that just isn't physically possible for most galaxies in areas of star formation. More general $\kappa$ which scale as powers of $\rho$ do not lead to realistic super-Eddington models either.

This left us with only the self-confinement model as a possible avenue for a cosmic ray Eddington limit. However, we found it still was unable to bring the Eddington star formation rate low enough to allow any of our galaxies to be realistically in reach of the Eddington limit. Thus, our analytical results suggested that a cosmic ray Eddington limit was not something that could be viably reached by any known galaxy. The dwarf starburst NGC 4449 comes within an order of magnitude of its Eddington limit, however, which suggests that cosmic ray blowout could play a role in limiting star formation in this type of galaxy, however - especially given the uncertainties in some of our parameters, and the simplifying assumptions made in constructing our models. We also found that almost all the cosmic ray energy injected at the base of our models has been expended as heat by the time the confinement radius is reached, and that the heating may be significant. Accounting for the effect of this heating is beyond the scope of this paper, but might contribute indirectly to an Eddington limit by raising the thermal gas pressure. 

To further our analysis, we continued with the self-confinement model but added in sources from star formation and losses from hadronic collisions. While keeping our form for collisions the same throughout our analysis, we modified the K-S Law to judge how a different star formation rate scaling with density would modify our findings. For one case, we assumed that $\alpha = 2$ in eqn. (\ref{eq:fulltransport}), pictured in Figure \ref{fig:CombinedUc0VaryPlotsSSL}, and in the other assumed that $\alpha = 1.4$, shown in Figure \ref{fig:CombinedUc0VaryPlotsSSL_LowerSFR}. 

As noted at the end of \S \ref{subsec:numResults}, the addition of sources and collisions to our pure streaming model does affect the Eddington limit for some of our galaxies. We find that collisions dominate over sources and streaming close to the inner boundary of our galaxies. However, our disks are very thin which leads to the collisions falling off rapidly and the streaming term dominating at larger radii. Therefore, the Eddington limit, especially for non-starburst galaxies gets pushed to larger values of $U_{c0}$. We also showed that the K-S Law one uses has a negligible effect on the Eddington limit. Since sources fall of so quickly with density compared to collisions and streaming, the K-S exponent we use will have largely no effect.

We also want to circle back to the main conclusions of both \cite{socrateseddington2008} and \cite{CrockerEddington2021Part1} to see how our conclusions compare to their own. In \cite{socrateseddington2008}, they theorized the existence of a cosmic ray Eddington limit, a point at which the cosmic ray energy density is so large that hydrostatic equilibrium is broken and the cosmic rays themselves drive a wind. We would argue our results seem to back up this conclusion. For most systems, typically the smaller galaxies we modeled, we found that a cosmic ray Eddington limit is reached. Our analysis indicates that cosmic ray streaming provides the best case scenario for reaching this Eddington limit but that once it is reached, the radius of confinement for the gas explodes and grows to large values well beyond the typical radius of a galaxy.

\cite{CrockerEddington2021Part1} found that, regardless of cosmic ray transport, as the gas column density was increased, cosmic rays became less important to the overall physics of the galaxy. Therefore, cosmic rays are only important in galaxies with low densities where collisional losses will be minimized. Our results seem to correlate well with this conclusion. We see that even with streaming, most of our systems (other than M82) would require extremely low gas densities on the order of $10^{-25} \unit g \unit cm^{-3}$ in order to break hydrostatic equilibrium. Furthermore, for three of our five galaxies, we see that when sources and hadronic collisions are added to our analysis, the Eddington limit becomes even more difficult to reach due to the strong calorimetry near the base of each galaxy. Therefore, the large gas densities utilized throughout our different galaxy models appears to be one of the main reasons why an Eddington limit is not reached in the parameter spaces of each galaxy. Although the particulars of our models and those of \cite{CrockerEddington2021Part2} are quite different, we find it reassuring that the galaxies we analyzed in common are found to be sub-Eddington within both their framework and ours (see Figure 4 of \cite{CrockerEddington2021Part2}).

Thus, we can see that an Eddington limit for cosmic rays does exist for most galaxies. However, regardless of the cosmic ray transport model used, the addition of sources and collisions, or the type of K-S Law used, we have found for our large range of models that no galaxy is realistically capable of reaching an Eddington limit for cosmic rays and launching a wind, based on the currently observed parameters. Our results seem to indicate that in systems with high density gas, the cosmic rays cannot build up enough of a pressure gradient to launch the wind, matching the results of \cite{CrockerEddington2021Part1}. However, based on the Eddington value for $\rho_\mathrm{0}$ in the streaming cases presented in Table \ref{tab:ana_densities}, it is possible that cosmic rays could move to regions of lower density gas in the galaxy and drive a wind. 

\acknowledgements
The authors would like to thank the referee for their thoughtful insights and suggestions that improved this paper. We would also like to thank Chad Bustard, Arianna Long, and Jay Gallagher for useful discussions. EMH and EGZ are supported by NSF AST 2007323 and the University of Wisconsin-Madison.

\appendix
\section{Non-Dimensional Equations}
\label{sec:nondimens}
In order to facilitate comparison between models, we non-dimensionalize all of our quantities first. If we assume spherical coordinates and simplify eqn. (\ref{eq:fulltransport}), we obtain:
\begin{equation}\label{eq:transport}
\begin{split}
    v_\mathrm{A}\frac{\partial P_\mathrm{c}}{\partial r} - \frac{\gamma_\mathrm{c} P_\mathrm{c} v_\mathrm{A}}{2\rho}\frac{\partial \rho}{\partial r} - \frac{1}{r^2}\frac{\partial}{\partial r}\lr{\kappa r^2 \frac{\partial P_\mathrm{c}}{\partial r}} \\
    = (\gamma - 1)\lr{S(\rho) - \frac{U_\mathrm{c}}{\tau_\mathrm{C}(\rho)}}
\end{split}
\end{equation}
where we have defined $Q = S(\rho) - U_\mathrm{c}/\tau_\mathrm{C}(\rho)$ where $S(\rho)$ is our sources term and $\tau_\mathrm{C}(\rho)$ represents the time between collisions for the cosmic-rays (hadronic loss time).

To make the equations nondimensional, we first define the new variables:
\begin{equation}
        x \equiv \frac{r}{R}  \hspace{1cm} p(x) \equiv \frac{P_\mathrm{c}(r)}{P_\mathrm{c0}} \hspace{1cm} s(x) \equiv \frac{\rho(r)}{\rho_\mathrm{0}}
\end{equation}
where we have non-dimensionalized $P_\mathrm{c}$ and $\rho$ by their base values at $r=R$. We define $R$ here as the inner radius at which we reach the boundary of our point mass which is a representation of the galaxy's core. 

For the Alfv\'{e}n speed, we know that $v_\mathrm{A} = B/(4\pi\rho_i)^{1/2}$. In order for $\bnabla \cdot \mathbf{B} = 0$, we must have $B \propto 1/r^2$. Therefore, since $v_\mathrm{A}$ is proportional to both $1/r^2$ and $\rho_i^{1/2}$, we can rewrite it as:
\begin{equation}
    v_\mathrm{A} = \frac{v_\mathrm{A,0}}{x^2s^{1/2}}
\end{equation}
where $v_\mathrm{A,0} = B_\mathrm{0}/(4\pi\rho_\mathrm{ion,0})^{1/2}$. Note that the Alfv\'{e}n speed is defined using the base ion density and not the total base density. However, we assume throughout this work that the plasma density and total gas density scale with $r$ in the same way.

We can similarly make the diffusion coefficient dimensionless by using $v_\mathrm{A,0}$ and defining:
\begin{equation}
    \chi = \frac{\kappa}{v_\mathrm{A,0}R} = \frac{\kappa\tau_\mathrm{A,0}}{R^2}
    \label{eq:dimens_diffusion_coeff}
\end{equation}
where we have defined a base Alfv\'{e}n transport time, $\tau_\mathrm{A,0} = R/v_\mathrm{A,0}$. Based on this definition, $\chi$ depends on the ion density as $\rho_\mathrm{ion,0}^{1/2}$ and will increase as the ion density increases. 

For the loss term, we will non-dimensionalize according to eqn. (13) from \cite{CrockerEddington2021Part1} which states:
\begin{equation}
    t_\mathrm{col} = 100\rho^{-1}_{-24} \rm Myr
    \label{eq:crocker_collisiontime}
\end{equation}
where $\rho_{-24} = \rho/10^{-24} \unit g \unit cm^{-3}$. Since the coefficient out in front may change based on the parameters of the galaxy modeled, we can more generally use a base value for the collisional loss time, $\tau_\mathrm{C,0}$, and define our loss time as:
\begin{equation}
   \tau_\mathrm{C} = \tau_\mathrm{C,0}\frac{\rho_0}{\rho} 
   \label{eq:collisiontime}
\end{equation}
where $\tau_\mathrm{C,0}\equiv\rm{100 Myr}(10^{-24}/\rho_0)$, and we have substituted in our dimensionless density as well. 

For the source term, since we already have a diffusion and loss time, it will be convenient to also have a base cosmic-ray injection time scale, $\tinjo$. The source term will also be proportional to the star formation rate which we take to be of the form $S \propto \rho^{\alpha}$ where $1<\alpha<2$ generally. Therefore, we can substitute for $S$:
\begin{equation}
    S = \frac{U_\mathrm{c0}s^{\alpha}}{\tinjo}
    \label{eq:injectiontime}
\end{equation}
where we have substituted in our dimensionless density. Equation (\ref{eq:injectiontime}) defines $\tinjo$.
 
Finally, we relate $P_c$ and $U_c$ by:
\begin{equation}
    P_\mathrm{c} = (\gamma - 1)U_\mathrm{c}
\label{eq:Pc_UcRelation}
\end{equation}
Substituting in these definitions and simplifying, eqn. (\ref{eq:transport}) becomes:
\begin{equation}
    s\frac{\partial p}{\partial x} - \frac{\gamma p}{2}\frac{\partial s}{\partial x} - s^{3/2}\frac{\partial}{\partial x}\lr{\chi x^2\frac{\partial p}{\partial x}} 
   = x^2s^{5/2}\lr{s^{\alpha - 1}c - p\ell}
    \label{eq:nondimens_transport}
\end{equation}
where we have defined $c = \tau_\mathrm{A,0}/\tinjo$ and $\ell = \tau_\mathrm{A,0}/\tau_\mathrm{C,0}$.

To get the non-dimensional form of the hydrostatic equilibrium equation, we 
rewrite eqn. (\ref{eq:hydroequil}) in terms of the scaled variables, yielding:
\begin{equation}
    \frac{P_\mathrm{c0}}{R}\frac{\partial p}{\partial x} = -\frac{\rho_\mathrm{0} s}{R} \frac{\partial \Phi}{\partial x} 
\label{eq:dimlesshse1}
\end{equation}

For the potential given by eqn. (\ref{eq:halomasspot}), we write:
\begin{equation}
\begin{split}
        \Phi &= -\frac{GM_{h}}{(r+a)} - \frac{GM_c}{r} \\
        &= - \frac{GM_c}{R}\lr{\frac{M_{h}}{M_c}\frac{1}{(x+a/R)} + \frac{1}{x}} \\
        &= \Phi_\mathrm{0}\lr{\frac{\mu}{(x+a/R)} + \frac{1}{x}}
\end{split}
\label{eq:dimlessphi}
\end{equation}
where we have defined $\mu = M_h/M_c$ to be the mass ratio between the halo and galaxy core
and have defined $\Phi_\mathrm{0} \equiv - GM_c/R$. Substituting eqn. (\ref{eq:dimlessphi}) into eqn. (\ref{eq:dimlesshse1}) yields:
\begin{equation}
    \frac{dp}{dx} = s\epsilon\lr{\frac{\mu}{(x+a/R)^2} + \frac{1}{x^2}}
    \label{eq:hydro_haloMass}
\end{equation}
where $\epsilon \equiv \Phi_\mathrm{0} \rho_\mathrm{0}/P_\mathrm{c0}$. 

Using Mathematica, we can then solve eqns. (\ref{eq:nondimens_transport}) and (\ref{eq:hydro_haloMass}) together and observe how both $p(x)$ and $s(x)$ change with respect to $x$. For these two potentials, we need to set some boundary conditions which we take to be $s(1) = 1$ and $p(1) = 1$. When diffusion is included, we need one more boundary condition, which is set by hydrostatic equilibrium and forces $p'(1) = \epsilon s(1)$.

For easier access, we put all of these different parameter values into Table \ref{tab:param_vals}. Note that these values will remain constant throughout all of our transport models and will remain unaltered by our choice of $\rho_\mathrm{0}$ and $U_\mathrm{c0}$.

In order to accurately solve these equations, we then need to find forms for the injection time and collision time. 
From eqn. (\ref{eq:crocker_collisiontime}), we can see that for $\rho_\mathrm{0} = 10^{-24} \unit g \unit cm^{-3}$, the collision time will be $100 \unit Myr$ which when converted to cgs units is $3.15 \times 10^{15}$ s. Since our collision time will vary inversely with the density, we find their product is:
\begin{equation}
    \tau_\mathrm{C,0}\rho_\mathrm{0} = (3.154 \times 10^{15})10^{-24} = 3.154 \times 10^{-9}
\label{eq:collision_constant}
\end{equation}
which we can use to find $\tau_\mathrm{C,0}$ for other galaxies.

We then similarly need to derive a value for $\tau_\mathrm{inj,0}$. From \cite{yoasthullstarbursts2016}, we know that the power imparted to cosmic-rays from supernovae is $7\times10^{48} \unit ergs/yr$ and that the volume of the CMZ is $2.5 \times 10^7 \unit pc^3$. Converting our units into CGS and finding the power per volume, we have:
\begin{equation}
    \frac{P_\mathrm{SN}}{V} = 3.02 \times 10^{-22} \unit \rm ergs \unit cm^{-3} \unit s^{-1} 
\end{equation}

From eqn. (\ref{eq:injectiontime}), we can see that this term has the same units as our source term, $S$. Therefore: 
\begin{equation}
    S = \frac{U_\mathrm{c0}s^{\alpha}}{\tinjo} = 3.02 \times 10^{-22} \unit ergs \unit cm^{-3} \unit s^{-1}.
\label{eq:sourceterm_M82}
\end{equation}

\begin{table}[]
    \centering
    \begin{tabular}{|c|c||c|c|}
    \hline
         $\mu$ & $M_h/M$  & $s$ & $\rho/\rho_\mathrm{0}$ \\
         $P_\mathrm{c0}$ & $U_\mathrm{c0}/3$ & $p$ & $P_\mathrm{c}/P_\mathrm{c0}$\\
         $\kappa$ & $3\times 10^{28} \unit cm^{2} \unit s^{-1}$ & $\chi$ & $\kappa\tau_\mathrm{A,0}/R^2$ \\
         $\Phi_\mathrm{0}$ & $GM/R$ & $\epsilon$ & $\Phi_\mathrm{0}\rho_\mathrm{0}/P_\mathrm{c0}$ \\
         $v_\mathrm{A,0}$ & $B_\mathrm{0}/(4\pi\rho_\mathrm{0})^{1/2}$ & $\tau_\mathrm{A,0}$ & $R/v_\mathrm{A,0}$ \\
         $c$ & $\tau_\mathrm{A,0}/\tinjo$ &  $\ell$ & $\tau_\mathrm{A,0}/\tlosso$ \\ 
         $x$ & $r/R$ & & \\
    \hline
    \end{tabular}
    \caption{Various Parameter Values and Variable Definitions for All Transport Models}
    \label{tab:param_vals}
\end{table}

For M82, $U_\mathrm{c0} = 525 \unit eV \unit cm^{-3}$ while at the base value of the density, $s=1$,
so eqn. (\ref{eq:injectiontime}) is satisfied for $\tinjo$:
\begin{equation}
    \tinjo = 2.78 \times 10^{12} \unit s = 8.82\times 10^{4} \unit yr
\end{equation}
 
The injection time for each galaxy will vary based on its density and the K-S Law that we used. Therefore, since we assume inside our mass distribution that star formation follows a K-S Law with $\alpha = 1.4$, we can use the product of $\tau_\mathrm{inj,0}$ and $\rho_\mathrm{0}^{1.4}$ to obtain a value for $\tau_\mathrm{inj,0}$ for other galaxies. For M82, the base value of the density can be taken to be $1.75\times10^{-21} \unit g \unit cm^{-3}$ and so the product becomes:
\begin{equation}\label{eq:tauninj0rho0}
    \tinjo\rho_\mathrm{0}^{1.4} = (2.78 \times 10^{12})(1.75\times10^{-21})^{1.4} = 2.42\times10^{-17}
\end{equation}
We can now use eqn. (\ref{eq:tauninj0rho0}) to find 
$\tau_\mathrm{inj,0}$ for any galaxy.

Finally, for convenience, we restate all of our non-dimensional equations here, along with the constants we solve for to keep the star formation and hadronic collision physics the same:
\begin{equation}
\begin{split}
    s\frac{\partial p}{\partial x} &- \frac{\gamma p}{2}\frac{\partial s}{\partial x} - s^{3/2}\frac{\partial}{\partial x}\lr{\chi x^2\frac{\partial p}{\partial x}}
   = x^2s^{5/2}\lr{s^{\alpha - 1}c - p\ell} \\
    &\frac{dp}{dx} = s\epsilon\lr{\frac{\mu}{(x+a/R)^2} + \frac{1}{x^2}} \\
   \tau_\mathrm{C,0}\rho_\mathrm{0} = 3.154 &\times 10^{-9} \hspace{2cm} \tinjo\rho_\mathrm{0}^{1.4} = 2.42\times10^{-17}
\end{split}
\label{eq:finaldimless}
\end{equation}
Figures \ref{fig:CombinedUc0VaryPlotsJS}, \ref{fig:CombinedUc0VaryPlotsSSL} and \ref{fig:CombinedUc0VaryPlotsSSL_LowerSFR}, which depict properties of the hydrostatic solutions for fixed based density $\rho_\mathrm{0}$ and varying $U_\mathrm{c0}$,  were made by repeatedly solving eqns. (\ref{eq:finaldimless}) for fixed values of $c$ and
$l$, $\chi\equiv 0$, and only varying $\epsilon$.


\begin{table*}[t!]
    \centering
    \begin{tabular}{|c||c|c|}
    \hline
         Galaxy & $c$ & $\ell$ \\
         \hline
         MW & $235.3$ & $412.7$ \\
         \hline
         M82 & $33.9$ & $52.2$ \\
         \hline
         LMC & $2.61$ & $30.3$ \\
         \hline
         NGC 4449 & $0.619$ & $8.11$ \\
         \hline
         DRC-8 & $911.7$ & $1008.9$ \\
    \hline
    \end{tabular}
    \caption{
    The calculated values for $c$ and $\ell$ for all five galaxies in our models. The definitions for $c$ and $\ell$ can be found in Table \ref{tab:param_vals}.}
    \label{tab:c_l_Values}
\end{table*}

Values of $c$ and $\ell$ are given  in Table \ref{tab:c_l_Values}
for the five galaxies modeled in this paper. We can see that for all galaxies, $\ell$ is much larger than $c$ which is expected since we find collisions dominate sources for all five systems.

\section{Galaxy Parameters}
\label{sec:galacticparam}
In this section, we outline the many different calculations we performed to obtain different parameter values throughout this paper, including the mass densities and cosmic ray energy densities. 

\subsection{Mass Densities}
For some galaxies, we are given the gas and ion number densities themselves and therefore just needed to multiply by a mean mass to the get the gas densities. However, for NGC 4449, we were given the the ion and gas mass for the galaxy and then converted that into a mass density by dividing it by the volume of the disk of that galaxy.

For the MW, we got the number densities from Figure 7 in \cite{FerriereISMGas2007} where $n_\mathrm{gas} \sim 10^{2.5} \unit cm^{-3}$ and $n_\mathrm{ion} \sim 10 \unit cm^{-3}.$ We then converted them to mass densities using eqn. (\ref{eq:massconvert}). We did the same thing for both M82 and the LMC. For M82, we got the number densities from Table 1 (molecular gas) and Table 3 (ionized gas) where $n_\mathrm{gas} \sim 400 \unit cm^{-3}$ and $n_\mathrm{ion} \sim 100 \unit cm^{-3}$. For the LMC, we got the number densities from \cite{BustardLMCWinds2020} of $n_\mathrm{gas} = n_\mathrm{ion} = 2 \unit cm^{-3}$.

For NGC 4449, we obtained the used the total gas surface densities from Table 2 of \cite{McQuinnDwarfStarburst2012} where they had that $\Sigma_\mathrm{gas} = 24.5 \unit M_{\odot} \unit pc^{-2}$. They assume that $\Sigma_\mathrm{ion} = 10 \unit M_{\odot} \unit pc^{-2}$ and so $\Sigma_{H_2} = 14.5 \unit M_{\odot} \unit pc^{-2}$. Converting to CGS units and then dividing by the height of the disk ($z= 200 \unit pc$), we obtain the mass densities listed in Table \ref{tab:galacticparams}.

For DRC-8, we estimated based on discussions with Arianna Long that $n_\mathrm{gas} = 10^3 \unit cm^{-3}$ and that $n_\mathrm{ion} = 10^{1.5} \unit cm^{-3}$. We then obtained the mass densities using the conversions from eqn. (\ref{eq:massconvert}).

\subsection{Cosmic Ray Energy Densities}

To calculate the base cosmic ray energy densities for NGC 4449 and DRC-8, we found the cosmic ray luminosity:
\begin{equation}
    L_\mathrm{c} = \frac{SFR\epsilon_\mathrm{c}}{m_\mathrm{SN}}
\end{equation}
where we took $\epsilon_\mathrm{c} = 10^{50} \unit ergs$ and $m_\mathrm{SN} = 100 \unit M_{\odot} \unit SN^{-1}$. 

We then found the cosmic ray lifetime using eqn. (\ref{eq:CRlifetime}) where:
\begin{equation}
    \tau_\mathrm{L} = \frac{\tau_T \tau_\mathrm{C}}{\tau_T+\tau_\mathrm{C}}
\end{equation}
where $\tau_T = z^2/\kappa$ (z is the height of the disk) and $\kappa = 3\times10^{28} \unit cm^2 \unit s^{-1}$) is the transport time and $\tau_\mathrm{C} = 3.2\times10^{-9}/\rho_\mathrm{0} \unit s$ is the collisional loss time.

We then get $U_\mathrm{c0}$ using eqn. (\ref{eq:Uc0fromSFR}) so that:
\begin{equation}
    U_\mathrm{c0} = \frac{L_\mathrm{c}\tau_\mathrm{L}}{V_{enc}}
\end{equation}
where $V_{enc}$ is the volume of the disk in which the star formation is occurring. For our solutions, $V_{enc} = \pi R^2 z$.

\bibliographystyle{aasjournal}
\bibliography{citations}

\begin{thebibliography}{}
\expandafter\ifx\csname natexlab\endcsname\relax\def\natexlab#1{#1}\fi
\providecommand{\url}[1]{\href{#1}{#1}}

\bibitem[{{Barger} {et~al.}(2016){Barger}, {Lehner}, \&
  {Howk}}]{BargerLMCWind2016}
{Barger}, K.~A., {Lehner}, N., \& {Howk}, J.~C. 2016, \apj, 817, 91

\bibitem[{{Booth} {et~al.}(2013){Booth}, {Agertz}, {Kravtsov}, \&
  {Gnedin}}]{boothwinds2013}
{Booth}, C.~M., {Agertz}, O., {Kravtsov}, A.~V., \& {Gnedin}, N.~Y. 2013,
  \apjl, 777, L16

\bibitem[{{Breitschwerdt} {et~al.}(1991){Breitschwerdt}, {McKenzie}, \&
  {Voelk}}]{breitschwerdtwinds1991}
{Breitschwerdt}, D., {McKenzie}, J.~F., \& {Voelk}, H.~J. 1991, \aap, 245, 79

\bibitem[{{Bustard} {et~al.}(2020){Bustard}, {Zweibel}, {D'Onghia},
  {Gallagher}, \& {Farber}}]{BustardLMCWinds2020}
{Bustard}, C., {Zweibel}, E.~G., {D'Onghia}, E., {Gallagher}, J.~S., I., \&
  {Farber}, R. 2020, \apj, 893, 29

\bibitem[{{Chy{\.z}y} {et~al.}(2000){Chy{\.z}y}, {Beck}, {Kohle}, {Klein}, \&
  {Urbanik}}]{ChyzyNGC4449_2000}
{Chy{\.z}y}, K.~T., {Beck}, R., {Kohle}, S., {Klein}, U., \& {Urbanik}, M.
  2000, \aap, 355, 128

\bibitem[{{Crocker} {et~al.}(2021{\natexlab{a}}){Crocker}, {Krumholz}, \&
  {Thompson}}]{CrockerEddington2021Part1}
{Crocker}, R.~M., {Krumholz}, M.~R., \& {Thompson}, T.~A. 2021{\natexlab{a}},
  \mnras, 502, 1312

\bibitem[{{Crocker} {et~al.}(2021{\natexlab{b}}){Crocker}, {Krumholz}, \&
  {Thompson}}]{CrockerEddington2021Part2}
---. 2021{\natexlab{b}}, \mnras, 503, 2651

\bibitem[{{Divakara Mayya} \& {Carrasco}(2009)}]{MayyaM82Summary2009}
{Divakara Mayya}, Y., \& {Carrasco}, L. 2009, arXiv e-prints, arXiv:0906.0757

\bibitem[{{Everett} {et~al.}(2008){Everett}, {Zweibel}, {Benjamin}, {McCammon},
  {Rocks}, \& {Gallagher}}]{everettwinds2008}
{Everett}, J.~E., {Zweibel}, E.~G., {Benjamin}, R.~A., {et~al.} 2008, \apj,
  674, 258

\bibitem[{{Ferri{\`e}re} {et~al.}(2007){Ferri{\`e}re}, {Gillard}, \&
  {Jean}}]{FerriereISMGas2007}
{Ferri{\`e}re}, K., {Gillard}, W., \& {Jean}, P. 2007, \aap, 467, 611

\bibitem[{{Girichidis} {et~al.}(2016){Girichidis}, {Naab}, {Walch}, {Hanasz},
  {Mac Low}, {Ostriker}, {Gatto}, {Peters}, {W{\"u}nsch}, {Glover}, {Klessen},
  {Clark}, \& {Baczynski}}]{girichidiswinds2016}
{Girichidis}, P., {Naab}, T., {Walch}, S., {et~al.} 2016, \apj, 816, L19

\bibitem[{{Guenduez} {et~al.}(2020){Guenduez}, {Becker Tjus}, {Ferri{\`e}re},
  \& {Dettmar}}]{GuenduezCMZMagnetic2020}
{Guenduez}, M., {Becker Tjus}, J., {Ferri{\`e}re}, K., \& {Dettmar}, R.~J.
  2020, \aap, 644, A71

\bibitem[{{Hanasz} {et~al.}(2013){Hanasz}, {Lesch}, {Naab}, {Gawryszczak},
  {Kowalik}, \& {W{\'o}lta{\'n}ski}}]{hanaszwinds2013}
{Hanasz}, M., {Lesch}, H., {Naab}, T., {et~al.} 2013, \apj, 777, L38

\bibitem[{{Hernquist}(1990)}]{HernquistPotential1990}
{Hernquist}, L. 1990, \apj, 356, 359

\bibitem[{{Hopkins} {et~al.}(2020){Hopkins}, {Chan}, {Garrison-Kimmel}, {Ji},
  {Su}, {Hummels}, {Kere{\v{s}}}, {Quataert}, \&
  {Faucher-Gigu{\`e}re}}]{HopkinsCRGalaxyForm2020}
{Hopkins}, P.~F., {Chan}, T.~K., {Garrison-Kimmel}, S., {et~al.} 2020, \mnras,
  492, 3465

\bibitem[{{Huang} \& {Davis}(2022)}]{HuangDavis2022}
{Huang}, X., \& {Davis}, S.~W. 2022, \mnras, arXiv:2105.11506

\bibitem[{{Ji} {et~al.}(2020){Ji}, {Chan}, {Hummels}, {Hopkins}, {Stern},
  {Kere{\v{s}}}, {Quataert}, {Faucher-Gigu{\`e}re}, \&
  {Murray}}]{JiCGMwithCRs2020}
{Ji}, S., {Chan}, T.~K., {Hummels}, C.~B., {et~al.} 2020, \mnras, 496, 4221

\bibitem[{{Kennicutt}(1998)}]{KennicuttSFR1998}
{Kennicutt}, Robert~C., J. 1998, \araa, 36, 189

\bibitem[{{Kennicutt} \& {De Los Reyes}(2021)}]{KennicuttSFRLaw2021}
{Kennicutt}, Robert~C., J., \& {De Los Reyes}, M. A.~C. 2021, \apj, 908, 61

\bibitem[{{Kulsrud} \& {Pearce}(1969)}]{KulsrudSelfConfine1969}
{Kulsrud}, R., \& {Pearce}, W.~P. 1969, \apj, 156, 445

\bibitem[{{Launhardt, R.} {et~al.}(2002){Launhardt, R.}, {Zylka, R.}, \&
  {Mezger, P. G.}}]{LaunhardtNBMassofMW2002}
{Launhardt, R.}, {Zylka, R.}, \& {Mezger, P. G.} 2002, A\&A, 384, 112

\bibitem[{{Liu} {et~al.}(2015){Liu}, {Gao}, \& {Greve}}]{LiuKSRevision2015}
{Liu}, L., {Gao}, Y., \& {Greve}, T.~R. 2015, \apj, 805, 31

\bibitem[{{Long} {et~al.}(2020){Long}, {Cooray}, {Ma}, {Casey}, {Wardlow},
  {Nayyeri}, {Ivison}, {Farrah}, \&
  {Dannerbauer}}]{LongYoungMassiveGalaxies2020}
{Long}, A.~S., {Cooray}, A., {Ma}, J., {et~al.} 2020, \apj, 898, 133

\bibitem[{{Lucchini} {et~al.}(2020){Lucchini}, {D'Onghia}, {Fox}, {Bustard},
  {Bland-Hawthorn}, \& {Zweibel}}]{LucchiniLMCStream2020}
{Lucchini}, S., {D'Onghia}, E., {Fox}, A.~J., {et~al.} 2020, \nat, 585, 203

\bibitem[{{Mannucci} {et~al.}(2005){Mannucci}, {Della Valle}, {Panagia},
  {Cappellaro}, {Cresci}, {Maiolino}, {Petrosian}, \&
  {Turatto}}]{MannucciSFRMassRate2005}
{Mannucci}, F., {Della Valle}, M., {Panagia}, N., {et~al.} 2005, \aap, 433, 807

\bibitem[{{Mao} \& {Ostriker}(2018)}]{MaoCRWind2018}
{Mao}, S.~A., \& {Ostriker}, E.~C. 2018, \apj, 854, 89

\bibitem[{{McQuinn} {et~al.}(2012){McQuinn}, {Skillman}, {Dalcanton},
  {Dolphin}, {Cannon}, {Holtzman}, {Weisz}, \&
  {Williams}}]{McQuinnDwarfStarburst2012}
{McQuinn}, K. B.~W., {Skillman}, E.~D., {Dalcanton}, J.~J., {et~al.} 2012,
  \apj, 751, 127

\bibitem[{{McQuinn} {et~al.}(2019){McQuinn}, {van Zee}, \&
  {Skillman}}]{McQuinnDwarfStarburst2019}
{McQuinn}, K. B.~W., {van Zee}, L., \& {Skillman}, E.~D. 2019, \apj, 886, 74

\bibitem[{{O'Connell} \& {Mangano}(1978)}]{OconnellM82Central1978}
{O'Connell}, R.~W., \& {Mangano}, J.~J. 1978, \apj, 221, 62

\bibitem[{{Oehm} {et~al.}(2017){Oehm}, {Thies}, \&
  {Kroupa}}]{OehmM81GroupEvol2017}
{Oehm}, W., {Thies}, I., \& {Kroupa}, P. 2017, \mnras, 467, 273

\bibitem[{{Parker}(1958)}]{ParkerSolarWind1958a}
{Parker}, E.~N. 1958, \apj, 128, 664

\bibitem[{{Posti} \& {Helmi}(2019)}]{PostiDMHaloMass2019}
{Posti}, L., \& {Helmi}, A. 2019, \aap, 621, A56

\bibitem[{{Ruszkowski} {et~al.}(2017){Ruszkowski}, {Yang}, \&
  {Zweibel}}]{ruszkowskiwinds2017}
{Ruszkowski}, M., {Yang}, H.-Y.~K., \& {Zweibel}, E. 2017, \apj, 834, 208

\bibitem[{{Salem} \& {Bryan}(2014)}]{SalemCROutflows2014}
{Salem}, M., \& {Bryan}, G.~L. 2014, \mnras, 437, 3312

\bibitem[{{Schmidt}(1959)}]{Schmidt59}
{Schmidt}, M. 1959, \apj, 129, 243

\bibitem[{{Simpson} {et~al.}(2016){Simpson}, {Pakmor}, {Marinacci}, {Pfrommer},
  {Springel}, {Glover}, {Clark}, \& {Smith}}]{simpsonWinds2016}
{Simpson}, C.~M., {Pakmor}, R., {Marinacci}, F., {et~al.} 2016, \apjl, 827, L29

\bibitem[{Socrates {et~al.}(2008)Socrates, Davis, \&
  Ramirez-Ruiz}]{socrateseddington2008}
Socrates, A., Davis, S.~W., \& Ramirez-Ruiz, E. 2008, \apj, 687, 202

\bibitem[{{Staveley-Smith} {et~al.}(2003){Staveley-Smith}, {Kim}, {Calabretta},
  {Haynes}, \& {Kesteven}}]{StaveleySmithLMCHI2003}
{Staveley-Smith}, L., {Kim}, S., {Calabretta}, M.~R., {Haynes}, R.~F., \&
  {Kesteven}, M.~J. 2003, \mnras, 339, 87

\bibitem[{{Strickland} \& {Heckman}(2009)}]{StricklandM82Wind2009}
{Strickland}, D.~K., \& {Heckman}, T.~M. 2009, \apj, 697, 2030

\bibitem[{{Strong} {et~al.}(2007){Strong}, {Moskalenko}, \&
  {Ptuskin}}]{StrongCRProp2007}
{Strong}, A.~W., {Moskalenko}, I.~V., \& {Ptuskin}, V.~S. 2007, Annual Review
  of Nuclear and Particle Science, 57, 285

\bibitem[{{Tumlinson} {et~al.}(2017){Tumlinson}, {Peeples}, \&
  {Werk}}]{TumlinsonCGM2017}
{Tumlinson}, J., {Peeples}, M.~S., \& {Werk}, J.~K. 2017, \araa, 55, 389

\bibitem[{{Uhlig} {et~al.}(2012){Uhlig}, {Pfrommer}, {Sharma}, {Nath},
  {En{\ss}lin}, \& {Springel}}]{uhligwinds2012}
{Uhlig}, M., {Pfrommer}, C., {Sharma}, M., {et~al.} 2012, \mnras, 423, 2374

\bibitem[{{Wiener} {et~al.}(2013){Wiener}, {Zweibel}, \& {Oh}}]{WienerWIM2013}
{Wiener}, J., {Zweibel}, E.~G., \& {Oh}, S.~P. 2013, \apj, 767, 87

\bibitem[{{Yoast-Hull} {et~al.}(2013){Yoast-Hull}, {Everett}, {Gallagher}, \&
  {Zweibel}}]{yoasthullm82_2013}
{Yoast-Hull}, T.~M., {Everett}, J.~E., {Gallagher}, J.~S., I., \& {Zweibel},
  E.~G. 2013, \apj, 768, 53

\bibitem[{{Yoast-Hull} {et~al.}(2014){Yoast-Hull}, {Gallagher}, \&
  {Zweibel}}]{yoasthullMWCMZ2014}
{Yoast-Hull}, T.~M., {Gallagher}, J.~S., I., \& {Zweibel}, E.~G. 2014, \apj,
  790, 86

\bibitem[{{Yoast-Hull} {et~al.}(2016){Yoast-Hull}, {Gallagher}, \&
  {Zweibel}}]{yoasthullstarbursts2016}
{Yoast-Hull}, T.~M., {Gallagher}, J.~S., \& {Zweibel}, E.~G. 2016, \mnras, 457,
  L29

\bibitem[{{Zweibel}(2017)}]{zweibelreview2017}
{Zweibel}, E.~G. 2017, Physics of Plasmas, 24, 055402

\end{thebibliography}

\end{document}